\documentclass[10pt,letterpaper]{article}
\usepackage[margin=0.7in]{geometry}
\usepackage{amsmath}
\usepackage{setspace}
\usepackage{amssymb}
\usepackage{amsthm}
\usepackage{amscd}
\usepackage{graphicx}
\usepackage{float}
\usepackage{cases}
\usepackage{subfig}
\usepackage{dsfont}
\usepackage{cite}
\usepackage{booktabs}
\usepackage[utf8]{inputenc}
\usepackage{wrapfig,framed,caption}
\usepackage{comment}

\newcommand\scalemath[2]{\scalebox{#1}{\mbox{\ensuremath{\displaystyle #2}}}}
\newtheorem{theorem}{Theorem}

\newtheorem{proposition}[theorem]{Proposition}

\DeclareGraphicsExtensions{.jpg}

\DeclareGraphicsExtensions{.eps}
\usepackage{epstopdf}
\usepackage{authblk}

\title{Dynamics of T Cell Receptor Distributions Following Acute Thymic Atrophy and Resumption}

\author[1]{Stephanie M. Lewkiewicz}
\author[2]{Yao-Li Chuang}
\author[3]{Tom Chou\thanks{Corresponding Author}}
\affil[1,3]{Department of Mathematics, UCLA, Los Angeles, CA,
  90095-1555, USA}
\affil[2]{Department of
  Mathematics, CalState Northridge, Northridge, CA 91330, USA}
\affil[3]{Department of Biomathematics, UCLA, Los
  Angeles, CA, 90095-1766, USA}

\begin{document}
\maketitle
\begin{abstract}
Naive human T cells are produced in the thymus, which atrophies
abruptly and severely in response to physical or psychological stress.
To understand how an instance of stress affects the size and
``diversity" of the peripheral naive T cell pool, we derive a
mean-field autonomous ODE model of T cell replenishment that allows us
to track the clone abundance distribution (the mean number of
different TCRs each represented by a specific number of cells).  We
identify equilibrium solutions that arise at different rates of T cell
production, and derive analytic approximations to the dominant
eigenvalues and eigenvectors of the problem linearized about these
equilibria.  From the forms of the eigenvalues and eigenvectors, we
estimate rates at which counts of clones of different sizes converge
to and depart from equilibrium values--that is, how the number of
clones of different sizes ``adjust'' to the changing rate of T cell
production.  Under most physiologically realistic realizations of our
model, the dominant eigenvalue (representing the slowest dynamics of
the clone abundance distribution) scales as a power law in the thymic
output for low output levels, but saturates at higher T cell
production rates. Our analysis provides a framework for quantitatively
understanding how the clone abundance distributions evolve under small
changes in the overall T cell production rate by the thymus.
\end{abstract}



\section{Introduction}
\label{sec:chapter2intro}

The thymus, a small organ located above the heart in humans, is a
crucial component of the primary lymphoid architecture, as the site of
T cell development~\cite{JANEWAY2012,GAMEIRO2010}.  The many different
T cell subpopulations together guide and assist the action of other
immune agents during infection~\cite{ALBERTS2002}, regulate the immune
response~\cite{CORTHAY2009}, and retain memory of encountered
pathogens~\cite{FARBER2014}.  As such, the thymus supplies the immune
compartment with its most essential source of direction, support, and
regulation.  T cells are produced when lymphocyte progenitors derived
from hematopoietic stem cells in the bone marrow migrate to the thymus
and begin a process of role selection, maturation, and vetting, before
being exported to the peripheral blood~\cite{TAKAHAMA2006}.  The most
significant event during thymocyte development is the rearrangement of
the $\alpha$ and $\beta$ chains of the T cell receptor
(TCR)~\cite{BASSING2002} that occurs in the thymic cortex which is
populated with thymocyte progenitors.  The particular rearrangement a
T cell undergoes determines its antigen specificity; a naive T cell in
the peripheral pool is activated when its TCR is bound by a cognate
antigen, a pathogen-derived peptide fragment capable of stimulating
that particular TCR~\cite{MASON1998}.  The total number of distinct
TCRs present across the full T cell pool is the ``TCR
diversity"~\cite{ZUGICH2004}, and this quantifies the breadth of the
pool's antigen responsiveness~\cite{LAYDON2015}.  Thymocytes also
undergo negative selection to eliminate cells that react too strongly
to self antigens presented by resident macrophages and dendritic
cells.  The small number of T cells that survive this process are
functionally competent and thus exported to the peripheral blood to
participate in the immune mechanism.

%

The thymus is known to experience both chronic and acute forms of
atrophy~\cite{CHAUDHRY2016}, resulting from both normal biological
processes and the presence of disease or stress. The most universal
form of thymic atrophy is age-related involution, the process by which
productive thymic tissue is gradually replaced with nonproductive
fat~\cite{STEINMANN1985}.  Involution begins at puberty and continues
indefinitely, and the resulting decline in T cell production has been
implicated as a likely source of immune dysfunction in the
elderly~\cite{GLOBERSON2000,GRUVER2007,SHANLEY2009}.  Acute atrophy
can occur under a plethora of conditions associated with a state of
disease or stress~\cite{GRUVER2008,DOOLEY2012,SELYE1936}, including
viral, bacterial, and fungal
infection~\cite{SAVINO2006,WANG1994,GRODY1985},
malnutrition~\cite{SAVINO2007}, cancer treatment~\cite{MACKALL1995},
bone marrow transplant~\cite{STOREK1995},
psychological stress, and
pregnancy~\cite{ZOLLER2007,TIBBETTS1999,RIJHSINGHANI1996}.  Each
condition facilitates thymic atrophy in (at least) one of several
ways, including reducing thymic cellularity~\cite{CHAUDHRY2016},
decreasing thymocyte proliferation and increasing
apoptosis~\cite{ASHWELL2000}, instigating premature export of
underdeveloped thymocytes~\cite{MENDES2006}, and inducing
morphological changes to the thymic
microenvironment~\cite{STANLEY1993}.  Such disturbances may
consequently alter the size and composition of the peripheral T cell
pool.  Decreased lymphocyte prevalence in the periphery during acute
involution has been
documented~\cite{GLAVINADURDOV2003,VANBAARLEN1989,JURETIC2001,FALKENBERG2014},
and \textit{Salmonella}, which infects the thymus itself, has been
shown to disrupt positive and negative selection, producing a skewed
TCR repertoire~\cite{LEVYARANGEL2015}.  Radiation and chemotherapy
drugs, such as temozolmide, used to treat cancer can also be highly
lymphotoxic, producing a lymphopenic state referred to as
``treatment-related lymphopenia"
(TRL)~\cite{YOVINO2013,MENDEZ2016,CAMPIAN2013}. Viral infections,
particularly HIV, and autoimmune disorders can induce lymphopenia by
increasing peripheral cellular death and redistributing cells to
inappropriate tissues, in addition to affecting production in the
thymus~\cite{LONG2012}. Congenital thymic aplasia, as seen in complete
DiGeorge syndrome, results in a lymphopenic state at birth
~\cite{CIUPE2009}.

The activation of the hypothalamic-pituitary-adrenal axis by stress
stimuli and subsequent release of glucocorticoids, which are known to
induce apoptosis in double-positive thymocytes~\cite{PURTON2004} and
inhibit their differentiation~\cite{ASHWELL2000}, is also likely a major
underlying catalyst of this acute
involution~\cite{DOOLEY2012,KONG2002}.  Evidence suggests that
glucocorticoid release is actually necessary to affect thymic
atrophy~\cite{ASHWELL2000,SELYE1936}.  Several other chemical agents
have been observed to participate in thymic atrophy, notably sex
hormones~\cite{DOOLEY2012}, which have been shown to weaken thymocyte
proliferation~\cite{ZOLLER2007} and induce
apoptosis~\cite{PATINO2000}, and the IL-6 cytokine family, which is
demonstrably thymosuppressive~\cite{GRUVER2008}.  Despite this
apparent sensitivity to stress, the thymus is highly plastic, and
generally recovers in size and functionality after removal of the
stressor~\cite{CHAUDHRY2016}.  Studies of the thymus during and after
chemotherapy treatment in cancer patients indicate a return of thymic
volume and productivity during recovery from
treatment~\cite{MACKALL1995}.  A recuperating thymus may even surpass
its pre-treatment volume, in a phenomenon known as ``thymic
rebound"~\cite{CHOYKE1987,COHEN1980}.  Such thymic recovery has also
been seen after infection~\cite{DEFRIEND2001} and traumatic
injury~\cite{GELFAND1972}. However, recovery is demonstrably
age-dependent, with thymi of older patients reconstituting the naive
T cell compartment more weakly than those of younger
patients~\cite{MACKALL1995}.  Although acute thymic atrophy has been
observed extensively in humans, much has yet to be learned about it,
and clear treatment protocol is lacking~\cite{CHAUDHRY2016}.

To this end, we present a mechanistic mathematical model to predict
changes in the size and diversity of the peripheral naive T cell
compartment in response to various immunologically diseased
conditions.  We study how this pool's size and composition adjust to
changes in the rate of thymic output.  We compartmentalize the
peripheral T cell pool by grouping clones--collections of T cells with
the same TCR--according to their size.  We then use a high-dimensional
autonomous ODE system to follow the time evolution of the number of
clones in each compartment.  We assume that the size of the peripheral
naive T cell pool is dictated by rates of thymic export of new
T cells, along with homeostatic proliferation and death mechanisms.
We assume a piecewise constant rate of thymic export, as the
atrophy/recovery cycle is known to be a rapid process, and that the
proliferative and death processes are subject to homeostatic
regulation based on the total T cell pool size.  We derive analytic
approximations to the dominant eigenvalues and eigenvectors of the
system linearized around its equilibria in both the presence and the
absence of thymic activity.  From this, we assess the rates of
convergence of different T cell compartments to equilibria that result
from a changing thymic export rate.  We then compare the linearized
and fully nonlinear models, and study several special cases.  We also
compute explicit representations of solutions in an
infinite-dimensional extension of our model.

\section{Mathematical Model and Analysis}
\label{sec:model}

We assume that the total naive T cell population  $N(t)$ in the immune
compartment (the blood and lymphatic tissue) satisfies a 
general ODE of the form

\begin{align}\label{eq:N!}
\frac{{\textrm d}N}{{\textrm d}t} = \gamma + p(N) N -\mu(N) N,
\end{align}
where $\gamma \geq 0$ is the rate of naive T cell export from the
thymus, and $p(N), \mu(N) \geq 0$ are regulated, $N$-dependent rates
of proliferation and death of naive T cells in the peripheral
bloodstream.  To prevent unbounded growth, we take $p(N)$ to be
non-increasing and $\mu(N)$ to be non-decreasing as cell counts, $N$,
increase.  We assume that $p(0) > \mu(0)$, as the lymphopenic
proliferation rate would be higher than the lymphopenic death
rate~\cite{BRADLEY2005,TAN2001,VIVIEN2001,FRY2005}.  At steady-state,
when a healthy, homeostatic cell count $N^*$ is achieved, $p(N^*) -
\mu(N^*) = -{\gamma \over N^{*}}\leq 0$. Note that when $\gamma = 0$,
simple decreasing functions $p(N)$ (and/or increasing functions
$\mu(N)$) admit multiple--typically two--fixed points. The $N=0$ fixed
point is unstable, while the one at $N>0$ is stable.

In order to compute the peripheral naive T cell diversity, we couple
Eq.~\ref{eq:N!}  with a system of ODEs that describes the time
evolution of the size-segregated subpopulations of the peripheral
naive T cell pool.  Let $c_k(t)$ denote the number of clones that are
of size $k$ at time $t \geq 0$. As formally shown in Song \& Chou
\cite{XU_JPA}, the \textit{mean} clone count is $c_{k} \propto P(n_{i}
= k,t)$, the marginalized probability that \textit{any} single clone
$i$ has population $k$. The master equation for $P(n_{i},t)$ is
difficult to solve with regulation terms.  Here, we provide a
heuristic derivation of the equations obeyed by $c_{k}(t)$ by using a
mean-field approximation that the \textit{total} population $N$ is
uncorrelated with any $n_{i}$ (although we know $N =
\sum_{i=1}^{\infty}n_{i}$).  Under this mean-field approximation, the
evolution of $P(n_{i} = k,t)$ and hence $c_{k}(t)$ obeys

\begin{align}
\frac{{\textrm d}c_1}{{\textrm d}t} &= \frac{\gamma}{\Omega} \left[ \Omega - \sum_{i=1}^M c_i - c_1\right] - p(N) c_1 + \mu(N)\left[2c_2 - c_1\right], \label{eq:CLONE1}\\
\frac{{\textrm d}c_k}{{\textrm d}t} &= \frac{\gamma}{\Omega}\left[c_{k-1}-c_k\right] + p(N)\left[(k-1)c_{k-1} - k c_k\right] + \mu(N) \left[(k+1)c_{k+1} - k c_k\right], \label{eq:CLONEk}\\
\frac{{\textrm d}c_M}{{\textrm d}t} &= \frac{\gamma}{\Omega} c_{M-1} + p(N) (M-1) c_{M-1} - \mu(N) M c_M, \label{eq:CLONEM}
\end{align}
where $k = 2, 3, \cdots, M-1$, $p(N)$ and $\mu(N)$ are approximated by
$p(N(t))$ and $\mu(N(t))$ (where $N(t)$ is the solution to
Eq.~\ref{eq:N!}), and the index $M$ in Eq.~\ref{eq:CLONEM} is the
hypothetical maximum size a clone can achieve.
%
%
We take $M$ to be finite for mathematical tractability and in
accordance with evidence of intraclonal competition that restricts
clone sizes and preserves a balanced TCR diversity~\cite{HATAYE2006}.
Each of the ODEs in Eqs.~\ref{eq:CLONE1}--\ref{eq:CLONEM} describes
how $c_k$ changes due to the effects of thymic export of new cells,
and proliferation and death in the periphery.  The constant $\Omega$
denotes the large total number of clonotypes that can potentially be
assembled in and exported from the thymus.  
%
%
The basic model includes immigration, birth, and death of multiple
species (\textit{i.e.}, clones) and can be developed in a fully
stochastic setting; however, in that case, only steady-state solutions
are available \cite{DESSALLES_JSP}.
%



In Eq.~\ref{eq:CLONEk}, the term $\frac{\gamma}{\Omega}$ represents
the rate at which cells of a given clonotype are exported to the
periphery from the thymus, and thus $\frac{\gamma}{\Omega} c_k$
represents the export rate of cells of clonotypes already represented
by size-$k$ clones in the periphery. The addition of a new cell to a
clone of population $k$ decreases by one the number of clones with
population $k$ and increases by one the number of clones with
population $k+1$.
%
Similarly, $\frac{\gamma}{\Omega} c_{k-1}$ represents the rate at
which clones move from $c_{k-1}$ to $c_k$ due to thymic export.  We
assume that cells immigrate, proliferate, or die one cell at a time,
forcing clones to move only among adjacent compartments. Thus, the
term $\frac{\gamma}{\Omega}\left[c_{k-1}-c_k\right]$ fully accounts
for changes to $c_k$ due to thymic export.
%
%
The term $p(N)k c_k$ denotes the rate at which cells in size-$k$
clones proliferate, which in turn corresponds to the rate at which
clones move from $c_k$ to $c_{k+1}$ due to peripheral proliferation.
Analogously, $p(N)(k-1)c_{k-1}$ denotes the rate at which
clones enter $c_k$ from $c_{k-1}$ due to
proliferation, so that $p(N)\left[(k-1)c_{k-1} - k c_k\right]$
accounts for changes to $c_k$ due to proliferation.  The death term in
Eq.~\ref{eq:CLONEk}, given by $\mu(N) \left[(k+1)c_{k+1} - k c_k
  \right]$ is defined analogously.  In Eqs.~\ref{eq:CLONE1}
and~\ref{eq:CLONEM}, we modify Eq.~\ref{eq:CLONEk} to account for
appropriate ``boundary conditions.''  In Eq.~\ref{eq:CLONE1}, the term
$\left[ \Omega - \sum_{i=1}^M c_i\right]$ gives the number of
clonotypes unrepresented in the periphery, so that
$\frac{\gamma}{\Omega} \left[\Omega-\sum_{i=1}^M c_i\right]$ provides
the rate at which new clones enter the periphery from the thymus.
Eq.~\ref{eq:CLONE1} also retains the terms from Eq.~\ref{eq:CLONEk}
that account for loss of clones in $c_1$ due to thymic export,
proliferation, and death, and the addition of clones into $c_1$ due to
death of cells in $c_2$.  Finally, Eq.~\ref{eq:CLONEM} retains terms
accounting for the introduction of clones into $c_M$ via thymic export
to and proliferation within clones in $c_{M-1}$, as well as loss of
clones from $c_M$ due to cellular death.  This represents a ``boundary
condition" that prevents clones from surpassing size $M$.

Summing Eqs.~\ref{eq:CLONE1}-\ref{eq:CLONEM}, we find that

\begin{align}\label{eq:N}
\frac{{\textrm d}\left(\sum_{k=1}^M k c_k\right)}{{\textrm d}t} 
= \gamma + p(N) \left(\sum_{k=1}^M k c_k\right) -
\mu(N) \left(\sum_{k=1}^M k c_k\right) - \left(p(N) M c_M + \frac{\gamma_0}{\Omega} c_M\right),
\end{align}

\noindent so that the ODE satisfied by $N(t)$ (Eq.~\ref{eq:N!}) and
that satisfied by $\sum_{k=1}^M k c_k(t)$ (Eq.~\ref{eq:N}) differ by
$\left(p(N) M c_M + \frac{\gamma_0}{\Omega} c_M\right)$, and thus
$N(t) \neq \sum_{k=1}^M k c_k$. Thus, very few numbers of clones of
large sizes $k > M$, whose population is accounted for in
Eq.~\ref{eq:N!}, are not accounted for fully in
Eqs.~\ref{eq:CLONE1}--\ref{eq:CLONEM}. This is especially salient in
the $\gamma = 0$ limit in which the $N>0$ fixed point is completely
``missed'' by the $c_{k}=0$ solution to
Eqs.~\ref{eq:CLONE1}--\ref{eq:CLONEM}. Thus, the $\gamma \to 0^{+}$
limit of the truncated system represents a singular limit where the
only solution to Eqs.~\ref{eq:CLONE1}--\ref{eq:CLONEM}, $c_{k}\to
0^{+}$, appears to violate the $N = \sum_{k=1}^{\infty}k c_{k} > 0$
constraint at the stable fixed point.


Nonetheless, we can use the ODE system
Eqs.~\ref{eq:CLONE1}--\ref{eq:CLONEM} to analyze the effects of
changes in the thymic output rate $\gamma$ provided we carefully use the
solutions of $c_{k}$ and $N$ that are consistent with the $N=0$ and
$N>0$ fixed points.
%
%
%
We denote the normal level of thymic export in an adult of a given age
by $\gamma = \gamma_0$.  To represent diminished thymic activity
during atrophy, we take $\gamma \ll \gamma_0$, or even $\gamma = 0$,
depending on the severity of the atrophy.  As the thymus is highly
plastic, the changes in $\gamma$ throughout the process of atrophy and
recovery tend to be rapid.  With this in mind, we model such cycles of
disease with a piecewise ODE system.  Specifically, let us observe a
human's response to disease-induced changes in thymic activity over
some time interval $I = [t_0, t_{S+1}]$.  We assume that this
individual's thymic export rate undergoes $S$ abrupt changes, at times
$t_1, t_2, \cdots, t_S$, where $t_0 < t_1 < t_2 < \ldots < t_S <
t_{S+1}$.  Letting $I_i = [t_i, t_{i+1}]$, so that $I =
\bigcup_{i=0}^S I_i$, we assume that $\gamma = \gamma_i \geq 0$, on
$I_i$.  If the initial condition $\{c_k(t_0)\}_{k=1}^M$ represents the
size of each $c_k$ compartment at the start of the process, we then
let $\{c_k^{i}(t)\}_{k=1}^M$ represent the solution of the ODE in
Eqs.~\ref{eq:N!}-\ref{eq:CLONEM}, on $I_i$, with $\gamma = \gamma_i$
and initial condition $\{c_k^i(t_i)\}_{k=1}^M =
\{c_k^{i-1}(t_i)\}_{k=1}^M$, for $i = 1, 2, \cdots, S$.  Thus, the
solution $\{c_k^{i}(t)\}_{k=1}^M$ represents the time evolution of the
$c_k$ compartments after a transition to a thymic activity level
$\gamma_i$.  This is the most general description of our model; in
practice, we will typically take $S=1$, representing a single abrupt
change in $\gamma(t)$.  Further descriptions of the piecewise ODE
formulation specific to certain disease patterns and particular
initial conditions are included in the relevant sections below.

Linear analysis of Eqs.~\ref{eq:CLONE1}--\ref{eq:CLONEM} will provide
information on how the \textit{clone counts} $c_{k}(t)$ evolve from
their steady-state values after a small perturbation in the system
(through changes in $\gamma$). The dynamics of $c_{k}$ are not
equivalent, but are qualitatively related to those of $n_{i}(t)$, the
number of cells in clone $i$. For example, when large values of
$n_{i}(t)$ increase, $c_{k\approx n_{i}}(t)$ decreases while
$c_{k\approx n_{i}+1}$ increases. Thus, increases in large $n_{i}$
convects $c_{k}$ forward, especially for larger $k$. As will be
explicitly shown, since $c_{k}$ is typically monotonically decreasing
in $k$, the dynamics of small(large) clone populations are correlated
with the dynamics of $c_{k}$ for small(large) $k$.

\section{Analysis for $\gamma > 0$ (functioning thymus)}
\label{sec:recovery}
We begin by studying the behavior of solutions of our ODE model under
the assumption of a strictly positive thymic export rate, $\gamma>0$.
We perform an analysis of equilibrium solutions of
Eqs.~\ref{eq:N!}-\ref{eq:CLONEM} and their stability, and also compute
an explicit solution in the infinite dimensional case that arises when
$M \to \infty$.  At the beginning of
sections~\ref{sec:recoveryinfinite} and~\ref{sec:recoveryfinite}
below, as well as in
sections~\ref{sec:atrophyanalyticsol},~\ref{sec:atrophylin},
and~\ref{sec:specialcases}, we focus on solutions over one individual
interval in the piecewise formulation described in
section~\ref{sec:model} above.  For simplicity when doing this, we
omit the $i$ notation that distinguishes the different subintervals,
writing $\gamma$ instead of $\gamma_i$, etc.  When the discussion
returns to the full piecewise ODE, the $i$ notation is reintroduced.

\subsection{Analytic Solution of the Infinite Dimensional System}
\label{sec:recoveryinfinite}
We begin by computing analytic expressions for the solutions $c_k$ of
Eqs.~\ref{eq:N!}-\ref{eq:CLONEM}.  If we take $M \to \infty$ and
consider instead the infinite dimensional system, the $c_k$
compartments can be obtained through a generating function in a
conjugate variable $q$, defined as

\begin{equation}
  Q(q,t) \equiv \sum_{k=0}^\infty c_k (t) q^k,
  \label{eq:GENERATING_FUNC}
\end{equation}
Note that $\left. \partial^k Q / \partial q^k \right|_{q = 0} = k!
c_k$ allows us to extract $c_k$ with $k \ge 0$.  In addition, the
total population can also be recovered
from the generating function via $\left. \partial Q/\partial q
\right|_{q = 1} = \sum_{k=0}^\infty k c_k = N$.

In order to derive an explicit form for $Q(q,t)$, we assume that an
explicit solution $N = N(t)$ of Eq.~\ref{eq:N!} can be found, so that
we may write $p$ and $\mu$ as functions of $t$ ($p = p(t)$, $\mu =
\mu(t)$).  By substituting Eqs.~\ref{eq:CLONE1},~\ref{eq:CLONEk} for
${\textrm d} c_k / {\textrm d}t$, the time derivative of $Q$ can be
expressed as

\begin{equation}
   \frac{\partial Q}{\partial t} = \sum_{k=0}^\infty
   \frac{{\textrm d} c_k}{{\textrm d}t} q^k = (q - 1)
(p(t)q - \mu(t)) \frac{\partial Q}{\partial q} + \frac{\gamma}{\Omega} (q-1) Q.
          \label{eq:DQDT0}
\end{equation}
The above partial differential equation can be solved analytically
along any characteristic curve $q(t)$ defined by the solutions to $\frac{{\textrm d}
  q}{{\textrm d} t} = - (q - 1) (p(t)q - \mu(t))$:

\begin{equation}
q(t) = 1+ {(1-q_{0})A(t) \over (1-q_{0})B(t) -1},
\end{equation}
where $q_0 = q(0)$ and

\begin{equation}
  A(t) \equiv \exp \left( - \int_0^t \left( p(s) - \mu(s) \right) {\textrm d}s \right)
\,\,\,\mbox{and}\,\,\, B(t) \equiv \int_0^t p(s) A(s) {\textrm d}s.
\label{eq:AB}
\end{equation}
Along each trajectory $q(t)$, the generating function
obeys $\frac{{\textrm d} Q}{{\textrm d} t} = -\frac{\gamma(t)}{\Omega} (1-q(t)) Q$
and can be expressed as

\begin{equation}
Q(q(t), t) = \sum_{k=0}^\infty c_k (0) q_0^k \exp \left(-\int_0^t
\frac{\gamma(s)}{\Omega} (1-q(s)) {\textrm d}s \right).
\label{eq:Q_NONZERO_GAMMA1}
\end{equation}
By allowing all possible initial values $q_{0}$, we can express the
full solution as

\begin{equation}
Q(q,t) = \sum_{k=0}^{\infty} c_{k}(0)\left[1-{1-q\over (1-q)B(t) +
    A(t)}\right]^{k} \exp\left[-\int_{0}^{t}\!\!\textrm{d} s {\gamma(s) \over
    \Omega}{(1-q)A(s) \over (1-q)(B(t)-B(s))+A(t)}\right].
\label{QSOLN}
\end{equation}
The solutions of $c_{k}(t)$ can then be extracted from $Q(q,t)$ by
taking a power expansion of $Q(q,t)$ and identifying coefficients with
$c_{k}(t)$ according to Eq.~\ref{eq:GENERATING_FUNC}. These exact
solutions to $c_{k}(t)$ will be compared with our subsequent results
derived from direct numerical solution of
Eqs.~\ref{eq:N!}--\ref{eq:CLONEM}.  By verifying that the time
evolution of $c_k$ under a finite dimensional formulation as in
Eqs.~\ref{eq:CLONE1}--\ref{eq:CLONEM} is sufficiently close to that of
$c_k$ under the infinite dimensional formulation in Eq.~\ref{QSOLN},
we allow the infinite and finite dimensional systems to be used more
or less interchangeably.  The finite dimensional formulation has the
advantage of not only admitting simple, explicit steady state
solutions, but also rates of convergence to steady state.

\subsection{Equilibrium Solution and Linearization}
\label{sec:recoveryfinite}

\noindent Returning to the truncated formulation (finite $M$), we now
study the equilibrium solution that results when taking $\gamma > 0$
in Eqs.~\ref{eq:N!}-\ref{eq:CLONEM}.  Denote such a generic
equilibrium solution by $\{c_k^*(\gamma)\}_{k=1}^M$, $N^*(\gamma)$.
For a given $N^*(\gamma)$, the $c_k^*(\gamma)$ have the form

\begin{align}
c_1^*(\gamma) &= \gamma\left[
\sum_{i = 1}^{M} \frac{\gamma/\Omega}{i! \mu(N^*(\gamma))^{i-1}}
\left(\prod_{j=1}^{i-1} \left[\frac{\gamma}{\Omega} + j
  p(N^*(\gamma))\right] \right)
+ \mu(N^*(\gamma))\right]^{-1},\label{eq:SS1aux}\\
c_k^*(\gamma) &= \frac{c_1^*(\gamma)}{k! \mu(N^*(\gamma))^{k-1}} 
\prod_{n=1}^{k-1}\left[\frac{\gamma}{\Omega} + n p(N^*(\gamma))\right].
\label{eq:SS2aux}
\end{align}

\noindent In the discussion below, we will write $N^*(\gamma)$ as
$N^*$ and $c_k^*(\gamma)$ as $c_k^*$ for simplicity, unless desiring
to emphasize the $\gamma$-dependence.  To identify the stability of
this equilibrium solution--and to identify the rates of convergence of
solutions to equilibria under the linearized model later on--we
consider the linearization of the system around this generic
equilibrium solution, represented by the $(M+1) \times (M+1)$ matrix
$L_S$ ($L_S = (s_{ij})_{1 \leq i,j \leq M+1}$), with
component $s_{ij}$ given by
\begin{align} \label{eq:S}
\begin{displaystyle}
  \left.
  \begin{cases}
    -\left(\frac{2 \gamma}{\Omega}\right) -(p(N^*)+\mu(N^*)), & \text{if } i = j = 1 \\
    -\left(\frac{\gamma}{\Omega}\right) + 2 \mu(N^*), & \text{if } i = 1, \text{  } j = 2 \\
    -\left(\frac{ \gamma}{\Omega}\right), & \text{if } i = 1; \text{  } 3 \leq j \leq M\\
    -\left(\frac{\gamma}{\Omega}\right) - i (p(N^*)+\mu(N^*)), & \text{if  } i = j; \text{  } 2 \leq j \leq M-1\\
 -\left(\frac{\gamma}{\Omega}\right) + i p(N^*), & \text{if } i = j+1; \text{ } 1 \leq j \leq M-1\\
    -M (p(N^*) + \mu(N^*)), & \text{if } i = j = M\\
    (i+1) \mu(N^*), & \text{if } i = j-1, \text{  } 2 \leq j \leq M\\
    p'(N^*)[(j-1) c_{j-1}^* - j c_j^*] + \mu'(N^*)[(j+1) c_{j+1}^* - j c_j^*], & \text{if } i = M+1; \text{  } 1 \leq j \leq M-1\\
    p'(N^*) (M-1) c_{M-1}^* - \mu'(N^*) M c_M^*, & \text{if } i = M+1; \text{  } j = M\\
    p'(N^*) N^* + p(N^*) - \mu'(N^*) N^* - \mu(N^*), & \text{if } i = j = M+1\\
    0, & \text{otherwise.  }
  \end{cases}
  \right\}
\end{displaystyle}
\end{align}

For clarity, an example of the matrix $L_S$ with $M = 4$ is
\begin{alignat*}{2}
\centering
\scalemath{0.725}{L_S =
\left(
  \begin{array}{ccccc}
    -\frac{2 \gamma}{\Omega} - (p(N^*)+\mu(N^*)) & -\frac{\gamma}{\Omega}
+ 2 \mu(N^*) & -\frac{\gamma}{\Omega} & -\frac{\gamma}{\Omega} & -p'(N^*) c_1^* \\[6mm]
    \frac{\gamma}{\Omega} + p(N^*) & -\frac{\gamma}{\Omega} - 2 (p(N^*)
+ \mu(N^*)) & 3 \mu(N^*) & 0 & p'(N^*) [c_1^* - 2 c_2^*]
+ \mu'(N^*) [3 c_3^* - 2 c_2^*] \\[6mm]
    0 & \frac{\gamma}{\Omega} + 2 p(N^*) & -\frac{\gamma}{\Omega}
- 3 (p(N^*) + \mu(N^*)) & 4 \mu(N^*) & p'(N^*)[2 c_2^* - 3 c_3^*] + \mu'(N^*)[4 c_4^* - 3 c_3^*] \\[6mm]
    0 & 0 & \frac{\gamma}{\Omega} + 3 p(N^*) & - 4\mu(N^*) & 3 p'(N^*) c_3^* - 4\mu(N^*) c_4^* \\[6mm]
    0 & 0 & 0 & 0 & p'(N^*) N^* + p(N^*) - \mu'(N^*) N^* - \mu(N^*) \\
  \end{array}
\right) }
\end{alignat*}
We now apply a simplifying assumption to the matrix $L_S$ to
analytically compute its eigenvalues more easily.  In general,
$\frac{\gamma}{\Omega} \sim 10^{-8}-10^{-6}$, and $p(N^*), \mu(N^*)
\sim 10^{-1}$ (when rates are measured in units of
$\text{year}^{-1}$), thus, we assume $\frac{\gamma}{\Omega} \ll
p(N^*), \mu(N^*)$ and define a new matrix $L_{\tilde{S}}$ which is
$L_{S}$ but with $\frac{\gamma}{\Omega}=0$ only in elements in which
it appears explicitly.  $L_{\tilde{S}}$ is defined by Eq.~\ref{eq:S}
with $\gamma/\Omega$ in the first five terms set to zero, but with
$N^{*}$ determined under the appropriate general immigration rate
$\gamma \geq 0$.

Numerical computation confirms that the eigenvalues of the new matrix
$L_{\tilde{S}}$ are essentially identical to those of the original
matrix $L_S({\gamma\over \Omega}\ll 1)$, validating our assumption
that the term $\frac{\gamma}{\Omega}$ may be neglected in $L_S$.  Note
that this assumption does not cause us to omit the constant $\gamma$
from the linearization matrix $L_S$ entirely, as the steady state
values $N^*$, $c_k^*$ depend on $\gamma$.  Denote by $\lambda_k^S$ for
$k = 1, 2, \cdots, M+1$ the eigenvalues of $L_{\tilde{S}}$, and note
that that the entry $\tilde{s}_{(M+1,M+1)} = \frac{{\textrm
    d}N}{{\textrm d}t}|_{N = N^*} = p'(N^*) N^* + p(N^*) - \mu'(N^*)
N^* - \mu(N^*)$ is an eigenvalue. We denote this eigenvalue by
$\lambda_{M+1}^S$.  Since $N^*$ is the stable equilibrium solution of
Eq.~\ref{eq:N!}, we assume that $\lambda_{M+1}^S < 0$.
%
%
The eigenvalues of $L_{\tilde{S}}$ all have strictly negative real
part, as do those of $L_S$.  The eigenvalues of the $(M+1) \times
(M+1)$ minor of $L_S$ can be approximated by $\tilde{\lambda}_k^S = k(p(N^{*}) -
\mu(N^{*}))$, with corresponding eigenvectors $\tilde{y}_k$ shown in the
following Proposition:

%
%

\begin{proposition} \label{prop:recovery1}
 If $p(N^*) - \mu(N^*) < 0$, the eigenvalues $\{\lambda_k^{S}\}_{k =
   1, \ldots, M}$ of the matrix $L_{\tilde{S}}$ are well approximated
 by the terms $\tilde{\lambda}_k^{S} = k(p(N^*) - \mu(N^*))$, in the
 sense that there exist vectors $\tilde{y}_k$ such that
 $||(L_{\tilde{S}} - \tilde{\lambda}_k^{S} I)\tilde{y}_k ||
 \longrightarrow 0$ as $M \longrightarrow \infty$.
\end{proposition}

\begin{proof}
We begin by assuming that the terms $\tilde{\lambda}_k^{S} = k (p(N^*)
- \mu(N^*))$ are themselves eigenvalues of $L_{\tilde{S}}$, and search
for their corresponding eigenvectors, $\tilde{y}_k = (\tilde{y}^1_k,
\tilde{y}^2_k, \cdots, \tilde{y}^M_k, 0)$. Choosing $\tilde{y}^1_k =
1$, we then choose $\tilde{y}^i_k$ for $i = 2, \ldots, M$ inductively
so as to force the $i$-th component of the residual vector, which we
denote by $[(L_{\tilde{S}} - \tilde{\lambda}_k^{S} I)\tilde{y}_k]_i$,
to equal zero for $i = 1, 2, \cdots, M-1$.  We then verify that
$[(L_{\tilde{S}} - \tilde{\lambda}_k^{S} I)\tilde{y}_k ]_{M}
\longrightarrow 0$ as $M \longrightarrow \infty$, so that for $M \gg
1$, $||(L_{\tilde{S}} - \tilde{\lambda}_k^{S} I)\tilde{y}_k|| \approx
0$, where $||\cdot ||$ is any $p$-norm.  (Trivially, $[(L_{\tilde{S}}
  - \tilde{\lambda}_k^S I)\tilde{y}_k]_{M+1} = 0$.)  We first note
that the components $\tilde{y}_k^1, \cdots, \tilde{y}_k^M$ of the
approximate eigenvector $\tilde{y}_k$ corresponding to eigenvalue
$\tilde{\lambda}_k^{S}$ are defined by the recurrence relation,

\begin{align} \label{eq:RECUR}
 i \tilde{y}^i_k = \left[(i + (k-1))\left(\frac{p(N^*)}{\mu(N^*)}\right)
+ (i-(k+1)) \right] \tilde{y}^{i-1}_k -
\left[i-2\right]\left(\frac{p(N^*)}{\mu(N^*)}\right) \tilde{y}^{i-2}_k.
\end{align}
The solution of this recurrence relation is then

\begin{align} \label{eq:RECURSOL}
\tilde{y}^i_k = \sum_{n=1}^{k} \frac{\left[ \prod_{j=1}^{n-1} (i-j)\right]
\left[ \prod_{j=1}^{k-n} (i+j)\right]}{k (-1)^{n-1} (n-1)! (k-n)!}
\left(\frac{p(N^*)}{\mu(N^*)}\right)^{i-n},
\end{align}
where we let $\prod_{j=1}^0 (i \pm j) = 1$, whenever such a
term appears in the above sum.  (This is verified in
Appendix~\ref{sec:APPrecursol}.)  As previously mentioned,
$[(L_{\tilde{S}} - \tilde{\lambda}_k^{S} I)\tilde{y}_k]_i = 0$ for $i
= 1, 2, \cdots, M-1$.  We now compute $[(L_{\tilde{S}} -
  \tilde{\lambda}_k^{S} I)\tilde{y}_k]_M$, obtaining

\begin{align}
[(L_{\tilde{S}} - \tilde{\lambda}_k^{S} I)\tilde{y}_k]_M &= (M-1) p(N^*)
\tilde{y}^{M-1}_k - M \mu(N^*) \tilde{y}^{M}_k \nonumber \\
 &= p(N^*) \sum_{n=1}^{k} \frac{\left[ \prod_{j=0}^{n-1} (M-1-j)\right]
\left[ \prod_{j=1}^{k-n} (M-1+j)\right]}{k (-1)^{n-1} (n-1)! (k-n)!}
\left(\frac{p(N^*)}{\mu(N^*)}\right)^{M-1-n} \nonumber \\
 & \hspace{1cm} - \mu(N^*) \sum_{n=1}^{k} \frac{\left[ \prod_{j=0}^{n-1} (M-j)\right]
\left[ \prod_{j=1}^{k-n} (M+j)\right]}{k (-1)^{n-1} (n-1)! (k-n)!}
\left(\frac{p(N^*)}{\mu(N^*)}\right)^{M-n}.
\label{eq:residual2}
\end{align}
Each term in the sum above has the form $p_k(M) a^M$, where
$p_k(M)$ is a polynomial of degree $k$ in the variable $M$, and $a =
p(N^*)/\mu(N^*)$.  Recalling that $p(N^*)/\mu(N^*) < 1$, then $\lim_{M
  \longrightarrow \infty} p_k(M) a^M = 0$, so that at large values of
$M$, $[(L_{\tilde{S}} - \tilde{\lambda}_k^{S} I) \tilde{y}_k ]_M
\approx 0$.  This demonstrates that $\tilde{\lambda}_k^{S}$ may be
regarded as an approximation to the true eigenvalue $\lambda_k^{S}$,
assuming that the eigenvalues of $L_{\tilde{S}}$ are stable under
small perturbations.
\end{proof}

\begin{figure}[h!]
\centering
\includegraphics[width=6.6in]{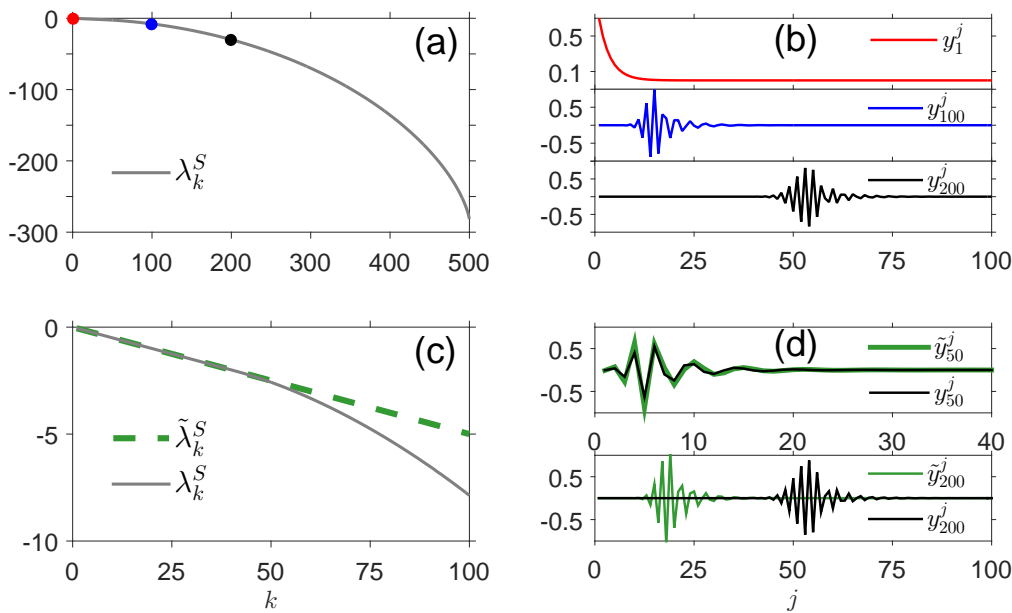}
\caption[Eigenvalues and eigenvectors of $L_{\tilde{S}}$, $\gamma >
  0$]{\textbf{Eigenvalues and eigenvectors of $L_{\tilde{S}}$, $\gamma
    > 0$.} Here, and in all subsequent evaluations, we parameterize
  our model using values qualitatively in a range based on human data,
  where the unit of rates are expressed in 1/year
  \cite{VRISEKOOP2008,DEBOER2013,WESTERA2015}. However, the model can
  have arbitrary units for general cases. (a) Numerically computed
  eigenvalue spectrum of the matrix $L_{\tilde{S}}$, with $p(N^*) =
  0.12$, $\mu(N^*) = 0.17$, and $M = 500$.  Dots identify the
  locations of the eigenvalues $\lambda_1^S$ (red), $\lambda_{100}^S$
  (blue), $\lambda_{200}^S$ (black).  (b) First 100 components
  ($\tilde{y}_k^1, \cdots, \tilde{y}_k^{100}$) of the eigenvectors
  with indices $k = 1, 100, 200$, the eigenvalues corresponding to
  which are marked on the spectral curve in (a).  (c) Comparison of
  true eigenvalues ($\lambda_k^S$) and approximate eigenvalues
  ($\tilde{\lambda}_k^S$) for $k = 1, \cdots, 100$.  Approximation is
  strong for $k \lesssim 50$. (d) (top) Comparison of
  $\tilde{y}_{50}^j$ and $y_{50}^j$ for $j = 1, \cdots, 40$, showing
  that the approximation is strong.  (bottom) Comparison of
  $\tilde{y}_{200}^j$ and $y_{200}^j$ for $j = 1, \cdots, 100$,
  showing that the accuracy of the approximation breaks down, but the
  qualitative behavior of $y_{200}^j$ is captured in
  $\tilde{y}_{200}^j$.}
\label{fig:thesis3fig1}
\end{figure}

The solutions $\tilde{y}_k^i$, as functions of $i$ for fixed $k$, are
characterized by patterns of oscillatory behavior, as shown in
Fig.~\ref{fig:thesis3fig1}, which depicts numerical solutions of true
and approximate eigenvalues and eigenvectors of $L_{\tilde{S}}$.  We
choose $p(N^*)$ and $\mu(N^*)$ to satisfy $p(N^*) < \mu(N^*)$, so that
at homeostatic population levels, the death rate exceeds the
proliferation rate, preventing exponential growth of the population.

Fig.~\ref{fig:thesis3fig1}(a) presents a sample plot of the eigenvalue
spectrum ${\lambda_k^S}$ in the case $M = 500$; on the spectral curve,
several eigenvalues are marked, for which the first $100$ components
of the corresponding eigenvectors are plotted
Fig.~\ref{fig:thesis3fig1}(b).  As discussed previously, all
eigenvalues are negative.  The eigenvector plots in
Fig.~\ref{fig:thesis3fig1}(b) indicate that as $k$ increases, the
oscillatory ``mass" of the corresponding eigenvectors occurs at
increasingly large values of $j$.  Fig.~\ref{fig:thesis3fig1}(c)
presents a comparison of the true eigenvalue spectrum ($\lambda_k^S$)
and the approximate eigenvalue spectrum ($\tilde{\lambda}_k^S$) of the
matrix $L_{\tilde{S}}$.  The eigenvalue approximation is very strong
for $1 \lesssim k \lesssim M/10$, and this remains true as $M$
varies.  The quantities $\tilde{\lambda}_k^S = k(p(N^*) - \mu(N^*))$
over-approximate the $\lambda_k^S$ for $k \gtrsim M/10$.
Fig.~\ref{fig:thesis3fig1}(d) depicts a comparison of $y_k$ and
$\tilde{y}_k$ for two values of $k$, one below the crossover $\sim
M/10$ ($k = 50$, top) and one above $M/10$ ($k = 200$, bottom).  As
expected, the eigenvector approximation is accurate precisely when
the corresponding eigenvalue approximation is accurate.  Even for $k
\gtrsim M/10$, the approximate eigenvectors $\tilde{y}_k$ present an
appearance similar to that of the $\tilde{y}_k$ for smaller $k$.  The
diminished accuracy of the eigenvalue/eigenvector pairs at higher $k$
(for fixed $M$) is attributable to the slower convergence of
$[(L_{\tilde{S}} - \tilde{\lambda}_k^S I)\tilde{y}_k]_M$ to $0$ as $M
\to \infty$ for larger $k$, which is immediately apparent from the
form in Eq.~\ref{eq:residual2}.  For a system of a fixed dimension
$M$, increasing $\mu(N^*)$ relative to $p(N^*)$ causes the approximate
eigenvalues $\tilde{\lambda}_k^S$ to become increasingly valid at
larger $k$, clearly due to the quicker convergence of the residual
quantity $||(L_{\tilde{S}} - \tilde{\lambda}_k^S I)\tilde{y}_k||$ to
$0$ when $p(N^*)/\mu(N^*) \ll 1$, as indicated by
Eq.~\ref{eq:RECURSOL}.



\begin{figure}[h]
\centering
\includegraphics[width=6.6in]{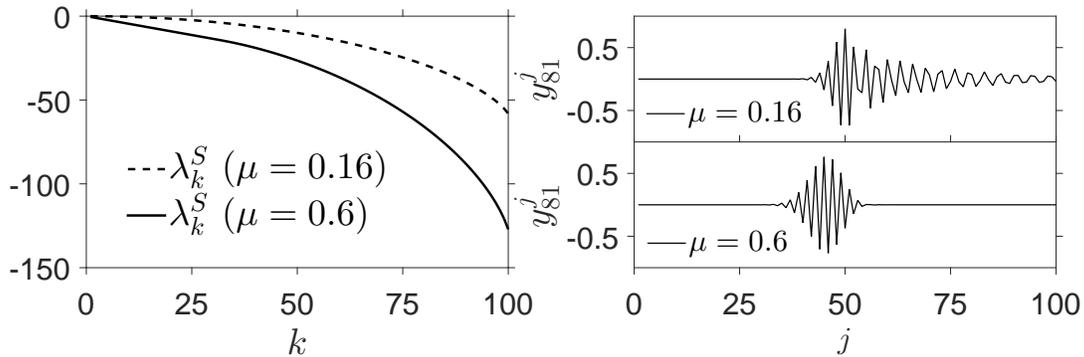}
    \caption[Eigenvalues and eigenvectors of $L_{\tilde{S}}$, $\gamma
      > 0$, varying $\mu$] {\textbf{Eigenvalues and eigenvectors of
        $L_{\tilde{S}}$, $\gamma > 0$, varying $\mu$.} (a)
      Numerically computed eigenvalues, $\lambda_k^S$, for $k = 1,
      \cdots, 100$, when $\mu(N^*) = 0.16$ and $\mu(N^*) = 0.6$.  In
      both cases, $p(N^*) = 0.15$, $M = 100$. (b) Numerically
      computed eigenvectors $y_{81}$.}
    \label{fig:VARYMU}
\end{figure}

Increases to $\mu(N^*)$ relative to $p(N^*)$ also cause an intensified
dampening of the oscillations in the eigenvector $\tilde{y}_k$ at
lower components $j$, which creates the illusion of the oscillatory
mass shifting to the left as $\mu(N^*)$ increases for fixed $p(N^*)$,
as in Fig.~\ref{fig:VARYMU}(b). At the same time, the entire
eigenvalue spectrum becomes more negative as $\mu(N^*)$ increases, as
indicated in Fig.~\ref{fig:VARYMU}(a), so that increases to the death
rate at homeostatic levels indicate much faster convergence to
equilibrium of all $c_k$ compartments.

\subsection{Behavior of the Linearized and Fully Nonlinear Systems}
\label{sec:recoveryanalysis}

This section addresses the convergence behavior of solutions in the
presence of a positive thymic export rate $\gamma > 0$.  This
situation represents a functioning thymus, with the possibility for
many different levels of functionality, ranging from total health
(high $\gamma \sim \gamma_0$) to dramatically diminished functionality
(low $\gamma$). It could also represent a new thymic export rate after
transplant of thymic tissue, as studied in the context of DiGeorge's
syndrome in Ciupe \textit{et al.}~\cite{CIUPE2009}.  In this case, we
determined that for each equilibrium solution of Eq.~\ref{eq:N!}, the
system has an equilibrium solution given by
Eqs.~\ref{eq:SS1aux},~\ref{eq:SS2aux}. If the steady state solution
$N^*(\gamma)>0$ of Eq.~\ref{eq:N!} represents a stable fixed point,
the corresponding equilibrium $c_k^*(\gamma)$ will also be stable
(typical regulated forms of proliferation and death, $p(N)$ and
$\mu(N)$, tend to result in one positive stable equilibrium solution
in Eq.~\ref{eq:N!}). If for some $i$, $\gamma_i > 0$, the solution
$\{c_k^i(t)\}_{k=1}^M$, $N^i(t)$ satisfies $c_k^i(t) \to
c_k^*(\gamma_i)$, $N^i(t) \to N^*(\gamma_i)$.  The convergence of the
total population $N^i(t) \to N^*$ occurs at rate $p'(N^*) N^* + p(N^*)
- \mu'(N^*) N^* - \mu(N^*)$.  Based on the eigenvalues and
eigenvectors of the approximate linearization, $L_{\tilde{S}}$, of
Eqs.~\ref{eq:N!}-\ref{eq:CLONEM} around this equilibrium, we can
formally construct approximations to the time-dependent solutions for
$c_{k}(t)$.

%
%

In the fully nonlinear system, however, the linearized eigenvalues
provide only \textit{a priori} rates of convergence of solution
trajectories initialized near equilibrium.  The accuracy of the
eigenvalues in providing convergence rates of solutions depends on the
initial conditions.  If the initial conditions, $c_k^i(t_i),
N^i(t_i)$, satisfy $c_k^i(t_i) \sim c_k^*(\gamma_i)$, $N^i(t_i) \sim
N^*(\gamma_i)$, then the solutions begin near the stable equilibrium,
and the eigenvalues provide accurate rates of convergence.  If the
initial conditions are far from equilibrium, the eigenvalues may not
provide accurate rates of convergence of the entire solution
trajectory.  When trajectories are far from equilibrium at time $t_i$,
further information about the speed of convergence can be discerned
from the relationship between $p(N^*(t_i)) - \mu(N^*(t_i))$ and
$p(N^*(\gamma_i)) - \mu(N^*(\gamma_i))$, the disparity in
proliferation and death rates at the starting and terminal population
levels.  If these quantities differ significantly, solution
trajectories are generally characterized by a transient period of fast
convergence, which carries the trajectory close enough to the stable
equilibrium that convergence rates from then on are those of the
linearized eigenvalues. For example, assume that an abrupt drop in
thymic productivity occurs at $t_i$, so that $\gamma_{i} <
\gamma_{i-1}$. As the steady-state naive T cell population evolves
from $N^{*}(\gamma_{i-1})$ to $N^{*}(\gamma_{i})$, for which $0 >
p(N^{*}(\gamma_i))-\mu(N^{*}(\gamma_i)) >
p(N^{*}(\gamma_{i-1}))-\mu(N^{*}(\gamma_{i-1}))$, the T cell pool will
experience a brief period of higher cellular death. As $N(t)$
approaches $N^{*}(\gamma_i)$, the convergence rates correspond to the
eigenvalues found from the linearized approximation.

\section{Analysis for $\gamma = 0$ (full thymic cessation)}
\label{sec:atrophy}
We now proceed to study the system after thymic export is shut off
($\gamma = 0$).  As in section~\ref{sec:recovery} above, we compute
equilibrium solutions of the truncated system (finite $M$) that arise
when $\gamma = 0$, and identify the rates of convergence of the
different $c_k(t)$ to equilibrium under the linearized model.  We also
take $M \to \infty$ and consider explicit solutions of the
infinite-dimensional system.

\subsection{Analytic Solutions}
\label{sec:atrophyanalyticsol}
In the $\gamma = 0$ case, the solution for $c_{k}(t)$ for $M\to
\infty$ can be readily expressed using the method of
characteristics. By using the generating function $Q(q,t)$ defined in
Eq.~\ref{QSOLN} and taking the $k$-th order derivative of $Q$ with
respect to $q$ at $q = 0$, we find

\begin{equation}
  c_k(t) = \left[ \frac{B(t)}{A(t) + B(t)} \right]^k
                \sum_{i=0}^\infty c_i (0)
\sum_{j = 0}^i {i \choose j} {k+j-1 \choose k}
           \left( 1 - \frac{1}{B(t)} \right)^{i-j}
           \left( \frac{A(t)}{B(t) \left( A(t) + B(t) \right)} \right)^j
           \label{eq:CK_GAMMA0}
\end{equation}
and $N(t) = A^{-1}(t) \sum_{k = 0}^\infty k c_k (0)$.  Note that for
depleted initial conditions $c_0 (0) = \Omega$ and $c_k (0) = 0$ for
$k \ge 1$, Eq.~\ref{eq:CK_GAMMA0} leads to $c_0 (t) = \Omega$ and $c_k
(t) = 0$ for $k \ge 1$ at all times.  Indeed, the T cell pool is
expected to remain empty since there is no thymic export.
%

\begin{figure}[h!]
\centering
 \includegraphics[width=5.6in]{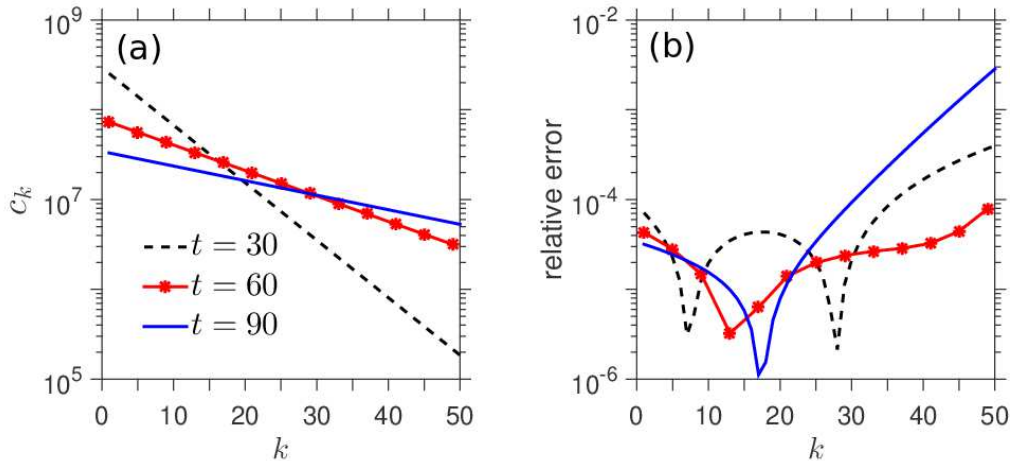}
    \caption[Computation of $c_k$ from method of characteristics,
      comparison with truncated system] {\textbf{Computation of $c_k$
        from method of characteristics, comparison with truncated
        system.} (a) Plots of $c_k$ for $k = 1, 2, \cdots, 50$, at
      times $t = 30, 60, 90$.  Solutions $c_k$ were computed
      numerically from the analytic method described
      in~\ref{sec:atrophyanalyticsol}, based on the
      infinite-dimensional system.  As a function of $k$, $c_k$
      presents as linear on a logarithmic scale.  (b) Relative error,
      $|c_k - y_k|/c_k$ for $k = 1, 2, \cdots, 50$, at times $t = 30$,
      $60$, $90$, where $c_k$ denotes the solutions depicted in (a),
      and $y_k$ denotes the numerically computed solutions of the
      truncated system in
      Eqs.~\ref{eq:CLONE1},~\ref{eq:CLONEk},~\ref{eq:CLONEM}.  From
      (b), the disparity between the solutions of the infinite
      dimensional systems ($c_k$) and the finite dimensional truncated
      systems ($y_k$) is negligible, validating our decision to use
      them interchangeably.  Coefficient functions: $p(N) = p_0 > 0$,
      $\mu(N) = \mu_0 + \mu_1(N^2/(N^2 + K^2))$.  Parameter values:
      $p_0 = 0.18$, $\mu_0 = 0.17$, $\mu_1 = 0.04$, $K = 10^{10}$,
      $\Omega = 10^{16}$, $M = 100$.  Initial condition: $c_0(0) =
      10^{16} - 10^{10}$, $c_1(0) = 10^{10}$, $c_k(0) = 0$ for $k \geq
      2$.}
    \label{fig:AnalyticSol}
\end{figure}

Fig~\ref{fig:AnalyticSol}(a) depicts a numerical computation of the
solution, $c_k$, of the infinite-dimensional formulation, obtained
from the generating function.  We include values of $c_k$ for $k = 1,
2, \cdots, 50$ at times $t = 30, 60, 90$.  As a function of $k$, the
$c_k$ present as linear on a logarithmic scale, as expected.  To
compare the infinite dimensional system with the truncated system, we
also compute solutions, $y_k$, of the truncated
Eqs.~\ref{eq:CLONE1}--\ref{eq:CLONEM} (not pictured).
Fig.~\ref{fig:AnalyticSol}(b) depicts the relative error, $|c_k -
y_k|/c_k$.  As we see, the error is several orders of magnitude
smaller than $c_k$, $y_k$ themselves at each of the times $t = 30, 60,
90$, indicating that the numerical solution of the truncated system
Eqs.~\ref{eq:CLONE1}-\ref{eq:CLONEM} is accurately described by the
exact method-of-characteristics solution and that the infinite- and
finite- dimensional systems may be used more or less interchangeably.

\subsection{Equilibrium Solutions and Linearization}
\label{sec:atrophylin}

We now investigate the solution to $c_{k}(t)$ near the fixed points
that arise when we take $\gamma = 0$ in
Eqs.~\ref{eq:N!}-\ref{eq:CLONEM}. In this case, the system has two
possible equilibrium solutions.  Denoting generic equilibrium
solutions by $\{c_k^*\}_{k=1}^M$ and $N^*$, the unstable solution is
$c_k^* = 0$ for all $k \geq 1$ and $N^* = 0$, and the asymptotically
stable solution is $c_k^* = 0$ for all $1\leq k \leq M$. However, in
the stable state we define $N^* = \tilde{N} > 0$, where $\tilde{N}$
satisfies $p(\tilde{N}) = \mu(\tilde{N})$.  To verify the stability of
these solutions, we consider the linearization of the system around
this equilibrium, which is represented by the $(M+1) \times (M+1)$
matrix we call $L_U$ ($L_U = (u_{ij})_{1 \leq i,j \leq M+1}$).  The
components $u_{ij}$ of $L_U$ are given explicitly by:

\begin{align} \label{eq:LU}
\begin{displaystyle}
 \left.
  \begin{cases}
    -j(p(N^*)+\mu(N^*)), & \text{if } i = j \leq M-1 \\
    -M \mu(N^*), & \text{if } i = j = M \\
    j \mu(N^*), & \text{if } i = j - 1; \text{  } 2 \leq j \leq M\\
    j p(N^*), & \text{if  } i = j + 1; \text{  } 1 \leq j \leq M-1\\
    p'(N^*)N^* + p(N^*) - \mu'(N^*)N^* - \mu(N^*), & \text{if  } i = j = M+1\\
    0. & \text{otherwise  }
 \end{cases}
 \right\}
\end{displaystyle}
\end{align}

\noindent As before, $u_{(M+1),(M+1)} = p'(N^*) N^* + p(N^*) -
\mu'(N^*) N^* - \mu(N^*)$ is an eigenvalue with eigenvector
$(0,\ldots,0,1)$, and the remaining eigenvalues are those of the
$(M+1) \times (M+1)$ minor of $L_U$.  As before, all eigenvalues of
the $(M+1) \times (M+1)$ minor have negative real part, and the
stability of an equilibrium solution depends on the sign of
$u_{(M+1),(M+1)}$.  If $N^* = 0$, then $u_{(M+1),(M+1)} = p(0) -
\mu(0) > 0$, as described previously, and the equilibrium $c_k^* = 0$,
$N^* = 0$ is unstable.  On the other hand, if $N^* = \tilde{N}$ with
$p(\tilde{N}) = \mu(\tilde{N})$, then $N^*$ represents a positive,
homeostatic cell count, and $u_{(M+1),(M+1)} = \left(p'(N^*) -
\mu'(N^*)\right)N^* < 0$, as $p(N), \mu(N)$ are assumed to be
non-increasing and non-decreasing, respectively.  Therefore, we have
that $c_k^* = 0$ and $N^* = \tilde{N}>0$ is a stable equilibrium
solution.

If $\gamma = 0$, the solution $\{c_k(t)\}_{k=1}^M$, $N(t)$, will
evolve away from the unstable equilibrium $c_k^* = 0$, $N^* = 0$ and
towards the equilibrium $c_k^* = 0$, $N^* = \tilde{N}>0$.  In this
instance, the pool of low-population clones is eradicated due
to lack of thymic productivity, and the high lymphopenic proliferation
rate pushes existent clones past the truncation threshold $M$, where
they are no longer accounted for in $c_k$ but are accounted for in
$N$, causing $N(t)\to N^*$ despite the fact that $c_k(t) \to 0$ for
all $k$.  As before, we wish to explore further the rates at which
individual functions $c_k$ diverge from the unstable fixed point
towards the stable one under the linearized and fully nonlinear
models.  To this end, we study the eigenvalues of the linearization
matrix, $L_U$, evaluated at the two equilibria.

First, consider the eigenvalues of $L_U$ evaluated at the unstable
equilibrium.  In this case, we assume $p(0) > \mu(0)$, as described
earlier.  Without thymic export, new clones are not generated in the
periphery, and existent clones expand due to the high proliferation
rate.  Under the dynamics described by
Eqs.~\ref{eq:CLONE1},~\ref{eq:CLONEk},~\ref{eq:CLONEM}, clones quickly
expand beyond the small-$k$ compartments and get ``caught" at the
boundary at size $M$, before depleting due to the slow death-induced
passage of single cell clones through the boundary at $k=1$.
According to Eq.~\ref{eq:N!}, the total cell population reaches a
natural homeostatic level through peripheral maintenance alone. To
investigate the rates at which these processes occur under the
linearized model, we derive approximations to the dominant eigenvalues
of $L_U$. Under the assumption that $p(0) > \mu(0)$, we denote the
true eigenvalues of $L_U$ by $\lambda_k^U$ for $k = 0, 1, \cdots M$,
with corresponding eigenvectors $z_k = (z_k^M, z_k^{M-1}, \cdots,
z_k^1, z_k^0)$ (note that we have reversed the index ordering here).
Assign to the eigenvalue $u_{(M+1),(M+1)} = p(0) - \mu(0)$ the label
$\lambda_k^M$, and to its eigenvector $(0,\ldots,0,1)$ the label
$z_M$. What remains is to find approximations to the other $M$
eigenvalues of $L_U$, which are precisely the eigenvalues of the
$(M+1) \times (M+1)$ minor.  For $i = 0, 1, \ldots, M-1$, denote the
approximation to the eigenvalue $\lambda_k^U$ by
$\tilde{\lambda}_k^U$, and the approximation to the eigenvector $z_k$
by $\tilde{z}_k = (\tilde{z}_k^M, \tilde{z}_k^{M-1}, \cdots,
\tilde{z}_k^1, 0)$.  We begin by establishing that the eigenvalue of
the $(M+1) \times (M+1)$ minor with the smallest magnitude,
$\lambda_0^U$, is well approximated by $\tilde{\lambda}_0^U = 0$.

\begin{proposition} \label{prop:recovery2}
The eigenvalue of $L_U$ of smallest magnitude, $\lambda_0^U$, is well
approximated by $\tilde{\lambda}_0^U = 0$, in the sense that there
exists a vector $\tilde{z}_0 = (\tilde{z}_0^M, \tilde{z}_0^{M-1},
\cdots, \tilde{z}_0^2, \tilde{z}_0^1, 0)$ such that $||(L_U -
\tilde{\lambda}_0^U I)\tilde{z}_0|| \longrightarrow 0$ as $M
\longrightarrow \infty$.
\end{proposition}

\begin{proof}
(Note: The components of $\tilde{z}_0$ are written above in
  ``descending" order for notational convenience, as this reflects the
  order in which they will be chosen recursively below.  The $i$-th
  component from the left of $\tilde{z}_0$, denoted explicitly by
  $\tilde{z}_0^{M-i+1}$, still corresponds to the function $c_i$.)  We
  begin by considering the matrix $(L_U - \tilde{\lambda}_0^U I) =
  L_U$ and searching for an appropriate eigenvector, $\tilde{z}_0$.
  We define $\tilde{z}_0^1 = 1$, and once again choose the components
  $\tilde{z}_0^i$ inductively via a three-term recurrence relation so
  as to force the $i$-th component of $(L_U - \tilde{\lambda}_0^U
  I)\tilde{z}_0$, which we denote as before by $[(L_U -
    \tilde{\lambda}_0^U I)\tilde{z}_0]_i$, to satisfy $[(L_U -
    \tilde{\lambda}_0^U I)\tilde{z}_0]_i = 0$ for $i = 2, 3, \cdots,
  M+1$.  While $[(L_U - \tilde{\lambda}_0^U I)\tilde{z}_0]_1 \neq 0$,
  we show that $[(L_U - \tilde{\lambda}_0^U I)\tilde{z}_0]_1
  \longrightarrow 0$ as $M \longrightarrow \infty$, so that
  $\tilde{z}_0$ may be regarded formally as an ``approximate"
  eigenvector corresponding to the approximate eigenvalue
  $\tilde{\lambda}_0^U$.

Defining $\tilde{z}_0^1 = 1$ and $\tilde{z}_0^2 = \frac{M
  \mu(0)}{(M-1)p(0)}$, we let $\tilde{z}_0$ be defined by solutions to
the recurrence relation,

\begin{align}
\tilde{z}_0^{i+2} = \left(\frac{(M-i) (\mu(0) + p(0))}{(M-(i+1))
  p(0)}\right) \tilde{z}_0^{i+1} -
\left(\frac{(M-(i-1))}{M-(i+1)}\right)\left(\frac{\mu(0)}{p(0)}\right)
\tilde{z}_0^i,
\end{align}
for $i = 1, 2, \cdots, M-2$.  It can be directly verified that the
solution to this recurrence relation is

\begin{align}
\tilde{z}_0^i = \left(\frac{M}{M-(i-1)}\right) \left(\frac{\mu(0)}{p(0)}\right)^{i-1}.
\end{align}
By construction of the recurrence relation, $[(L_U -
  \tilde{\lambda}_0^{U} I)\tilde{z}_0]_i = 0$ for $i = 2, 3, \cdots, M$.
The first component, $[(L_U - \tilde{\lambda}_0^{U} I)\tilde{z}_0]_1$,
satisfies

\begin{align}
[(L_U - \tilde{\lambda}_0^{U} I)\tilde{z}_0]_1 =
-\mu(0) M \left(\frac{\mu(0)}{p(0)}\right)^{M-1} \longrightarrow 0
\end{align}
as $M \longrightarrow \infty$.  Thus, when $M \gg 1$,
$||(L_U - \tilde{\lambda}_0^U I)\tilde{z}_0|| \approx 0$, and we may
conclude that $\tilde{\lambda}_0^U = 0$ and $\tilde{z}_0$ are suitable
approximations to $\lambda_0^U$ and $z_0$, respectively.
\end{proof}

Recalling that all eigenvalues of $L_U$ have negative real parts, the
true eigenvalue $\lambda_0^U$ has a negative real part of very small
magnitude.  We now identify which entries in the eigenvector $z_0$, as
approximated by $\tilde{z}_0$, are particularly large in magnitude in
comparison with the others.  Recalling that the $i$-th component of
$\tilde{z}_0$ is given by $\tilde{z}_0^i = (M/(M-(i-1)))
(\mu(0)/p(0))^{i-1}$, the $\tilde{z}_0^i$ decay nearly exponentially
in $i$, so that $c_k$ for large $k$ are preserved by this slow
eigenvalue, in agreement with the described ``build up" of clones at
the boundary $k = M$ when $\gamma = 0$.

The eigenvalue $\lambda_1^U$ of second smallest magnitude is well
separated from $\lambda_0^U$, and it encodes information about how the
number of small clones evolve. Similar analysis of an eigenvalue
$\tilde{\lambda}_1^U$ and eigenvector $\tilde{z}_1$ approximating
$\lambda_1^U$ and $z_1$ indicates that $c_k$ empties much more rapidly
for small $k$ than it does for large $k$, as small clones expand in
size and race to the boundary at $k = M$.  In particular, all $c_k$
except those with $k \sim M$, which had been preserved by the slow
eigenvalue $\lambda_0^U$, empty at nearly the same rate, $\lambda_1^U
\approx \tilde{\lambda}_1^U = (\mu(0) - p(0))$.

\begin{proposition} \label{prop:recovery3}
In the case $p(0) > \mu(0)$, the matrix $L_U$ has an eigenvalue,
$\lambda_1^U$, which is well approximated by $\tilde{\lambda}_1^U =
(\mu(0) - p(0))$, in the sense that there exists a vector
$\tilde{z}_1$ such that $||(L_U - \tilde{\lambda}_1^U I)\tilde{z}_1||
\longrightarrow 0$ as $M \longrightarrow \infty$.
\end{proposition}

\begin{proof}
First define $\tilde{z}_1^1 = 1$, and $\tilde{z}_1^2 =
\left(\frac{M+1}{M-1}\right)\left(\frac{\mu(0)}{p(0)}\right) -
\frac{1}{M-1}$.  Then for $i = 1, 2, \cdots, M-2$, let
$\tilde{z}_0^{i+2}$ be given by the solutions to the following
recurrence relation:

\begin{align} \label{eq:RECUR1}
\tilde{z}_1^{i+2} = \tilde{z}_1^{i+1} + \left(\frac{M-(i-1)}{M-(i+1)}\right) \left(\frac{\mu(0)}{p(0)}\right) \left(\tilde{z}_1^{i+1} - \tilde{z}_1^i\right).
\end{align}
It is worth nothing that if we were to instead choose
$\tilde{z}_1^1 = \tilde{z}_1^2$, then the recurrence relation in
Eq.~\ref{eq:RECUR1} would have a constant solution, $\tilde{z}_1^i =
\tilde{z}_1^1$ for all $i = 1, 2, \cdots, M$.  Although $\tilde{z}_1^1
\neq \tilde{z}_1^2$ for our purposes, solutions of the recurrence
relation do converge rapidly to constants, as will be discussed
later.

By construction, $[(L_U - \tilde{\lambda}_1^U I)\tilde{z}_1]_i = 0 $
for $ i = 2, 3, \cdots, M+1$.  Additionally, $[(L_U -
  \tilde{\lambda}_1^U I)\tilde{z}_1]_1 = 2\mu(0)(\tilde{z}_1^M -
\tilde{z}_1^{M-1})$.  To show that $[(L_U - \tilde{\lambda}_1^U
  I)\tilde{z}_1]_1 \longrightarrow 0$ as $M \longrightarrow \infty$,
we use Eq.~\ref{eq:RECUR1} to derive an explicit bound on the quantity
$2\mu(0)(\tilde{z}_1^M - \tilde{z}_1^{M-1})$. Consider

\begin{align}
|\tilde{z}_1^{i+2} - \tilde{z}_1^{i+1}| &= \left(\frac{M-(i-1)}{M-(i+1)}\right) \left(\frac{\mu(0)}{p(0)}\right) |\tilde{z}_1^{i+1} - \tilde{z}_1^i| \nonumber \\
&= \left(\frac{M-(i-1)}{M-(i+1)}\right) \left(\frac{M-(i-2)}{M-i}\right) \left(\frac{\mu(0)}{p(0)}\right)^2 |\tilde{z}_1^i - \tilde{z}_1^{i-1}| \nonumber \\
%
%
& \: \hspace{1cm} \vdots \nonumber \\
&= \left(\frac{M-(i-(j-1))}{M - (i+1)}\right)\left(\frac{M-(i-j)}{M-i}\right) \left(\frac{\mu(0)}{p(0)}\right)^j |\tilde{z}_1^{i+2-j} - \tilde{z}_1^{i+1-j}| \nonumber \\
& \: \hspace{1cm} \vdots \nonumber \\
&= \left(\frac{M-1}{M-(i+1)}\right) \left(\frac{M}{M-i}\right) \left(\frac{\mu(0)}{p(0)}\right)^i |\tilde{z}_1^2 - \tilde{z}_1^1|. \label{eq:BOUND}
\end{align}
Taking $i = M-2$ in the above relation, we find

\begin{align}
|[(L_U - \tilde{\lambda}_1^U I)\tilde{z}_1]|_{1} &= 2\mu(0)|\tilde{z}_1^M - \tilde{z}_1^{M-1}| \nonumber \\
&= \mu(0) \left(M(M-1)\right)\left(\frac{\mu(0)}{p(0)}\right)^{M-2}
|\tilde{z}_1^2 - \tilde{z}_1^1| \longrightarrow 0 \nonumber
\end{align}
as $M \longrightarrow \infty$.  Thus, for $M \gg 1$, $||(L_U -
\tilde{\lambda}_1^U I)\tilde{z}_1|| \approx 0$, and we find that
$\tilde{z}_1$ is ``almost" an eigenvector of $L_U$ corresponding to
the approximate eigenvalue $\tilde{\lambda}_1^U$.
\end{proof}

We now wish to identify which of the $c_k$ empty at the rate
determined by the second approximate eigenvalue,
$\tilde{\lambda}_1^U$.  As it turns out, the eigenvector $\tilde{z}_1$
corresponding to this eigenvalue is ``nearly" constant, and thus all
$c_k$ empty at essentially the same rate.  We see this by identifying
that even though we are only concerned with a finite number ($M$) of
terms of the sequence generated by the recurrence relation in
Eq.~\ref{eq:RECUR1}, as the index $M$ becomes infinitely large, the
sequence $\{z_1^i\}_{i=1}^M$ exhibits ``Cauchy-like" behavior,
mimicking ``convergence" to a limiting value.  We make this more
precise in the following Proposition:

\begin{proposition} \label{prop:recovery4}
Let $\tilde{z}_1$ be an approximate eigenvector of $L_U$ corresponding
to approximate eigenvalue $\tilde{\lambda}_1^U$, where $\tilde{z}_1$
is generated by the recurrence relation in Eq.~\ref{eq:RECUR1}.  Then
at large $M$, the components of $\tilde{z}_1$ exhibit ``Cauchy-like"
behavior: for any $\varepsilon > 0$ and $0 < c < 1$, we may choose $M
\in \mathbf{N}$ such that $|\tilde{z}_1^{m} - \tilde{z}_1^{n}| <
\varepsilon$ for all $cM \leq m \leq M$ and $cM \leq n \leq M$.
\end{proposition}

\begin{proof}
Recalling the bound on $|\tilde{z}_1^{i+2} - \tilde{z}_1^{i+1}|$
obtained in Eq.~\ref{eq:BOUND}, we find

\begin{align*}
|\tilde{z}_1^m - \tilde{z}_1^n| &= \left| \sum_{i=n}^{m-1} \left( \tilde{z}_1^{i+1} - \tilde{z}_1^i\right)\right| \leq \sum_{i = n}^{m-1} |\tilde{z}_1^{i+1} - \tilde{z}_1^i|\\
&= \sum_{i=n}^{m-1} \left(\frac{M-1}{M-i}\right)\left(\frac{M}{M-(i-1)}\right)
\left(\frac{\mu(0)}{p(0)}\right)^{i-1} |\tilde{z}_1^2 - \tilde{z}_1^1|\\
&\leq M(M-1) |\tilde{z}_1^2 - \tilde{z}_1^1| \sum_{i=n}^{m-1}
\left(\frac{\mu(0)}{p(0)}\right)^{i-1}\\
&= M(M-1) |\tilde{z}_1^2 - \tilde{z}_1^1|
\left(\frac{\mu(0)}{p(0)}\right)^{n-1}
\left(\sum_{j=0}^{m-n-1} \left(\frac{\mu(0)}{p(0)}\right)^j\right)\\
&= M(M-1)|\tilde{z}_1^2 - \tilde{z}_1^1| \left(\frac{\mu(0)}{p(0)}\right)^{n-1}
\left(\frac{1 - \left(\frac{\mu(0)}{p(0)}\right)^{m-n}}{1 - \frac{\mu(0)}{p(0)}}\right)\\
&\leq \frac{|\tilde{z}_1^2 - \tilde{z}_1^1|}{\left(1 - \frac{\mu(0)}{p(0)}\right)}
M(M-1)\left(\frac{\mu(0)}{p(0)}\right)^{n-1}
\leq \frac{|\tilde{z}_1^2 - \tilde{z}_1^1|}
%
%
{\left(1 - \frac{\mu(0)}{p(0)}\right)}
\left[M(M-1) \left(\frac{\mu(0)}{p(0)}\right)^{cM-1}\right]. \\
%
%
%
\end{align*}
Recalling that $M(M-1)\left(\frac{\mu(0)}{p(0)}\right)^{cM}
\longrightarrow 0$ as $M \longrightarrow \infty$, we may choose $M$
large enough such that

\begin{align}
M(M-1)\left(\frac{\mu(0)}{p(0)}\right)^{cM} \leq \varepsilon
\frac{\left[1-\left({\mu(0)\over p(0)}\right)\right]
\left({\mu(0)\over p(0)}\right)}{|\tilde{z}_1^2 - \tilde{z}_1^1|},
%
\end{align}
allowing us to conclude that $|\tilde{z}_1^m - \tilde{z}_1^n| <
\varepsilon$ for all $M \geq m, n \geq cM$.  \end{proof}

By taking $0 < c, \varepsilon \ll 1$, we find that ``most" components
of $\tilde{z}_1$ are within an $\varepsilon$-distance of each other,
so that the eigenvector is nearly constant.  (The constancy breaks
down at the components representing large-$k$ compartments.)  With
this, we are able to classify the rates at which all of the $c_k$ are
lost from the pool.  As the eigenvector corresponding to the second
smallest magnitude eigenvalue, $\lambda_1^{U} \approx
\tilde{\lambda}_1^{U}$, is nearly constant, all $c_k$, except those
with $k \sim M$, empty at nearly the same rate, on a time scale $\sim
|\mu(0) - p(0)|^{-1}$.

\begin{figure}[h!]
\centering
\includegraphics[width=6.5in]{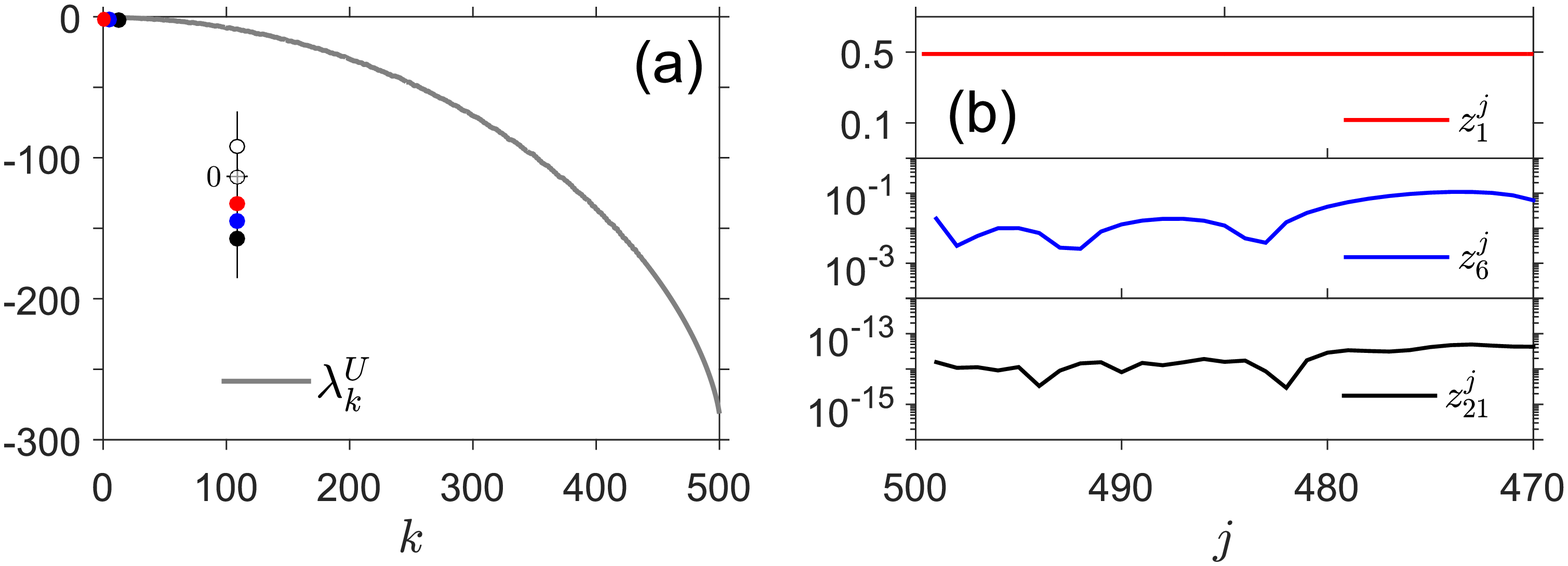}
\caption[Eigenvalues and eigenvectors of $L_{\tilde{S}}$, $\gamma =
  0$]{\textbf{Eigenvalues and eigenvectors of $L_{\tilde{S}}$, $\gamma
    = 0$.} (a) Numerically computed eigenvalue spectrum of the matrix
  $L_{U}$, with $p(0) = 0.17$, $\mu(0) = 0.12$, and $M = 500$.  Dots
  identify the locations of the eigenvalues $\lambda_1^U$ (red),
  $\lambda_{6}^U$ (blue), $\lambda_{21}^U$ (black). The inset shows
  the eigenvalues of the full system, including the positive
  eigenvalue (when $p(0)>\mu(0)$) and the eigenvalue
  $\lambda_{0}^{U}\approx 0$.  (b) First 30 components ($z_k^{500},
  \cdots, z_k^{470}$) of the eigenvectors with indices $k = 1, 6, 21$,
  the eigenvalues corresponding to which are marked on the spectral
  curve in (a).  (Note that the eigenvectors were defined in a
  ``reverse-order", so that $z_k^{500}$ corresponds to compartment
  $c_1$, $z_k^{495}$ to compartment $c_6$, and $z_k^{480}$ to
  compartment $c_{21}$ (generally: $z_k^{M-j+1}$ corresponds to
  compartment $c_j$).}
\label{fig:thesis3fig3}
\end{figure}

The remaining eigenvalues $\lambda_2^U, \cdots, \lambda_{M-1}^U$ are
not treated analytically, but numerical computation indicates that a
similar general approximation to the $k$-th eigenvalue,
$\tilde{\lambda}_k^U$, may be made, taking the form
$\tilde{\lambda}_k^U = k(\mu(0) - p(0))$.  The true eigenvalues,
$\lambda_k^U$, are depicted in Fig.~\ref{fig:thesis3fig3}(a)
(Comparison of the true and approximate spectra is omitted, as the
result is similar to that depicted in Fig.~\ref{fig:thesis3fig1}(c).
That is, $\lambda_k^U \approx k(\mu(0) - p(0))$ if $k \lesssim
M/10$.).  The oscillatory behavior observed in the approximate
eigenvectors in the case $p(N^*) < \mu(N^*)$ of
section~\ref{sec:recovery} is absent here; although the subsequent
approximate eigenvectors $\tilde{z}_2, \cdots, \tilde{z}_{M-1}$ do not
share the Cauchy-like behavior of $\tilde{z}_1$, the components
corresponding to $c_k$ for small $k$ do not vary much in magnitude,
and are thus interpreted as being nearly constant themselves
(Fig.~\ref{fig:thesis3fig3}(b).)  Thus, we conclude that for small
$k$, the functions $c_k$ all have very similar dynamics, diverging at
the rate $|\mu(0) - p(0)|$, while for large $k$, the $c_k$ converge
very slowly, at a rate governed by the dominant, near-zero eigenvalue
$\lambda_0^U$.

We now consider the stable equilibrium solution, $c_{k\leq M}^* = 0$ for $k
\geq 1$, $N^* = \tilde{N} > 0$.  In this case, the linearization
around this equilibrium may be expressed as $p(N^*) L_{U'}$, where
$L_{U'} = (u'_{ij})_{1\leq i,j \leq M+1}$ is the $(M+1) \times (M+1)$
matrix with component $u'_{ij}$ given by

\begin{align} \label{eq:LU'}
\begin{displaystyle}
 \left.
  \begin{cases}
    -2j, & \text{if } i = j \leq M-1 \\
    -M, & \text{if } i = j = M \\
    j, & \text{if } i = j - 1; \text{  } 2 \leq j \leq M\\
    j, & \text{if  } i = j + 1; \text{  } 1 \leq j \leq M-1\\
    \frac{p'(N^*)N^* + p(N^*) - \mu'(N^*)N^* - \mu(N^*)}{p(N^*)}, & \text{if  } i = j = M+1\\
    0. & \text{otherwise  }
 \end{cases}
 \right\}
\end{displaystyle}
\end{align}

\noindent For reference, an example of the matrix $L_{U'}$, with $M =
4$, is given below:
\begin{alignat}{2} \label{eq:LU'exp}
\centering
\left(
  \begin{array}{ccccc}
    - 2 & 2 & 0 & 0 &  0 \\[6mm]
    1 & - 4 & 3  & 0 & 0 \\[6mm]
    0 & 2  & - 6 & 4 &  0 \\[6mm]
    0 & 0 & 3  & - 4  & 0 \\[6mm]
    0 & 0 & 0 & 0 & \frac{p'(N^*) N^* + p(N^*) - \mu'(N^*) N^* - \mu(N^*)}{p(N^*)}  \\
  \end{array}
\right)
\end{alignat}

Denote by $\lambda_k^{U'}$ for $k = 1, 2, \cdots, M, M+1$ the
eigenvalues of the matrix $L_{U'}$ evaluated at the stable equilibrium
solution, and $x_k = (x_k^1, x_k^2, \cdots, x_k^M, x_k^{M+1})$ their
corresponding eigenvectors.  As before, let $\lambda_{M+1}^{U'} =
{1\over p(N^{*})}{\partial [(p-\mu)N]\over \partial N}\vert_{N=N^{*}}
<0$.
%
%
Then, the remaining eigenvalues $\lambda_1^{U'}, \cdots,
\lambda_M^{U'}$ are those of the $(M+1) \times (M+1)$ minor of
$L_{U'}$, which are independent of the parameters of the system except
$M$.  (Of course, the eigenvalues of $p(N^*) L_{U'}$ are then $p(N^*)
\lambda_k^{U'}$ for $k = 1, \cdots, M+1$.)  These eigenvalues and
eigenvectors are not treated analytically.  The analysis conducted in
section~\ref{sec:recovery} does not apply, as it relied on the
assumption $p(N^*) < \mu(N^*)$, which no longer holds.  However,
numerical computation indicates that in the case $p(N^*) = \mu(N^*)$,
which applies here, the eigenvalue spectrum and associated
eigenvectors qualitatively resemble the $\lambda_k^S$, $y_k$ studied
analytically in section~\ref{sec:recovery}.  


\subsection{Behavior of the Linearized and Fully Nonlinear Systems}
\label{sec:atrophyanalysis}
We now interpret these results in the context of the particular
diseased states to which they naturally apply.  We first identified an
unstable equilibrium solution, $c_k^* = 0$, $N^* = 0$, and studied the
linearization of the system around this equilibrium.  Under the
linearized model, if $\gamma_i = 0$ for some $i$, the
eigenvalue/eigenvectors pairs suggest that solutions diverge away from
this equilibrium, with $c_k^i$ for small $k$ evolving at a rate $\sim
\lambda_1^U = (\mu(0) - p(0))$, and $c_k^i$ for $k \sim M$ evolving at
the very slow rate given by the small-magnitude eigenvalue,
$\lambda_0^U$.  The total population $N^i(t)$ evolves at the rate
$(p(0) - \mu(0))$.  This situation represents the repopulation of the
T cell pool from a small number of cells via peripheral proliferation
within a highly pathological state involving both complete thymic
inactivity (e.g. thymectomy or total functional cessation) and near
full lymphopenia (as may result from treatment regimens for cancer,
etc.).

We then identified a stable equilibrium solution, $c_k^* = 0$, $N^* =
\tilde{N} > 0$.  As $\tilde{N}$ is asymptotically stable, $N^i(t) \to
\tilde{N}$ after diverging from $N^* = 0$.  Under the linearized
model, the eigenvalue/eigenvector pairs suggest that $c_k^i(t) \to 0$
slowly for small $k$, and $c_k^i(t) \to 0$ much more quickly for large
$k$.

As before, the validity of the eigenvalues in providing accurate
convergence rates of $c_k^i, N^i$ to and from equilibria depends on
the initial condition $c_k^i(t_i)$ in the full nonlinear model.  If
the human is in a state of immune health for $t < t_i$, so that
$\gamma_{i-1} > 0$ and the initial conditions $c_k^i(t_i), N^i(t_i) >
0$ satisfy $c_k^i(t_i) \sim c_k^*(\gamma_{i-1})$, $N^i(t_i) \sim
N^*(\gamma_{i-1})$, we expect that $p(N^i(t_i)) < \mu(N^i(t_i))$.  The
higher rate of death than proliferation at $t_i$ may cause a transient
period of quick collapse, with $N^i(t)$ decreasing to $\tilde{N}$.  As
$N^i(t) \to \tilde{N}$, convergence occurs at the rates dictated by
the linearized eigenvalues.  If $\gamma_{i-1} > 0$ but $c_k^i(t_i),
N^i(t_i) \sim 0$, so that the thymus is functioning to some extent but
the T cell pool has been eradicated, trajectories first diverge away
from the unstable zero equilibrium at rates given by the linearized
eigenvalues.  As $p(N^i(t_i)) - \mu(N^i(t_i)) \to p(\tilde{N}) -
\mu(\tilde{N}) = 0$, the motion of trajectories transitions from being
dictated by the eigenvalues of the unstable equilibrium to those of
the stable equilibrium.

In summary, we show in Proposition \ref{prop:recovery2} that $c_k$ of
larger $k$ is sustained by a near-zero eigenvalue $\lambda_{0}^{U}
\simeq 0$, as the solution evolves away from the unstable empty state
when $\gamma = 0$. Furthermore, in Propositions \ref{prop:recovery3}
and \ref{prop:recovery4} we identify a series of negative eigenvalues,
and a uniform eigenvector corresponding to the slowest decay rate
$\lambda_{1}^{U} < 0$.  It suggests a uniform asymptotic decay of
components other than the larger $k$ components preserved by
$\lambda_{0}^{U}$.

\section{Special Cases and Numerical Evaluation}
\label{sec:specialcases}

Using the approximate rates of convergence provided by linearization,
we can now study the time scale of the T cell pool's adjustment to a
new export rate.  While some T cell clones will expand and attain a
large size, most are small.  Thus, in both the cases of thymic atrophy
and recovery, we take as a proxy for the rate at which the T cell
diversity converges to equilibrium the eigenvalue that dictates the
rate of convergence of $c_1$, typically given by the quantity $p(N^*)
- \mu(N^*)$ (this tends to also be the dominant eigenvalue).
Recalling that $p(N^*) < \mu(N^*)$, high proliferation rates ($p(N^*)
\sim \mu(N^*)$) lead to small values of $|p(N^*) - \mu(N^*)|$, thus
slower adjustment of diversity to the changing $\gamma$.  That is, in
a proliferation-dominant scenario, a drop in thymic export leads to
repopulation via clonal expansion. In this section, we study several
specific models arising from canonical choices of $p$ and $\mu$, and
compute the changing convergence rate as gamma varies.

\subsection{The Logistic Model: regulated proliferation, constant death}
\label{sec:logisticmodel}

\begin{figure}[h]
\centering
\includegraphics[width=6.5in]{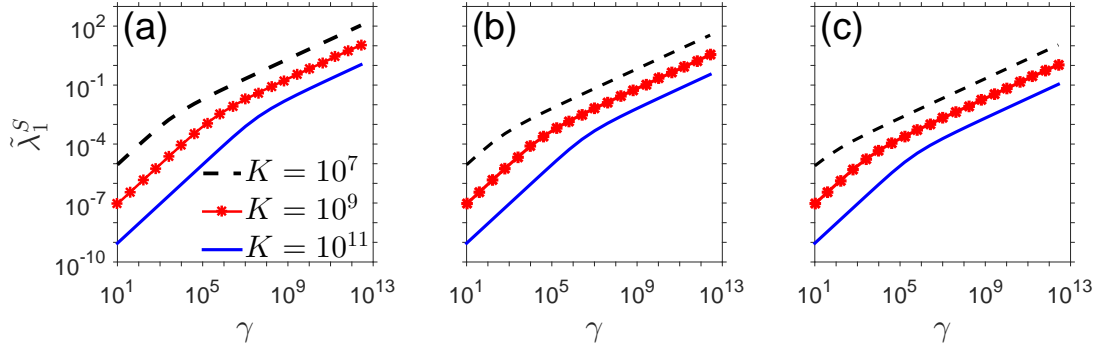}
    \caption[Dominant eigenvalue, $\tilde{\lambda}_1^S$, of
      $L_{\tilde{S}}$, plotted against $\gamma$, Case
      1]{\textbf{Dominant eigenvalue, $\tilde{\lambda}_1^S$, of
        $L_{\tilde{S}}$, plotted against $\gamma$, Case 1.} In (a),
      $p_0 = 0.18$, $\mu_0 = 0.17$.  In (b), $p_0 = 0.018$, $\mu_0 =
      0.017$.  In (c), $p_0 = 0.0018$, $\mu_0 = 0.0017$.  There is an
      approximate power law relationship between $\tilde{\lambda}_1^S$
      and $\gamma$ within this range of parameter values.  In (a), for
      example, the best fit line to the curve $K = 10^7$ is given by
      $\log(\tilde{\lambda}_1^S) = (2.673\times 10^{-11}) \log(\gamma)
      + 5.594$, with $R^2 = 0.8904$.  The curve $K = 10^9$ is fitted
      by $\log(\tilde{\lambda}_1^S) = (2.672 \times 10^{-12})
      \log(\gamma) + 0.558$, with $R^2 = 0.8905$, and the curve $K =
      10^{11}$ is fitted by $\log(\tilde{\lambda}_1^S) = (2.67 \times
      10^{-13}) \log(\gamma) + 0.05478$, with $R^2 = 0.8911$.  }
    \label{fig:Case1}
\end{figure}

We begin with the canonical logistic growth model, taking $p(N) = p_0
(1 - N/K)$, $\mu(N) = \mu_0$, where $p_0, \mu_0 > 0$ are basal rates
of cellular proliferation and death, respectively, and $K > 0$ is an
inherent carrying capacity.  Under this model, Eq.~\ref{eq:N!} has a
positive steady state, $N^*$, given by

\begin{align} \label{eq:SSN}
N^* = \left(\frac{K}{2 p_0}\right) \left((p_0 - \mu_0)
+ \sqrt{(p_0 - \mu_0)^2 + \frac{4 \gamma p_0}{K}}\right).
\end{align}
In this case, $p(N^*) - \mu(N^*) = p_0 \left(1 - \frac{N^*}{K}\right)
- \mu_0 = \frac{1}{2}\left((p_0 - \mu_0) - \sqrt{(p_0 -
  \mu_0)^2 + \frac{4 \gamma p_0}{K}}\right) < 0$, so that the
assumption $p(N^*) < \mu(N^*)$ always applies.  Moreover,
$\tilde{\lambda}^S_{M+1} = - \sqrt{(p_0 - \mu_0)^2 + \frac{4 \gamma
    p_0}{K}}$, so it is clear that $0 > \tilde{\lambda}^S_1 >
\tilde{\lambda}^S_{M+1}$, and $\tilde{\lambda}^S_1$ is the dominant
eigenvalue.  Then,

\begin{align} \label{eq:logisticeval}
|p(N^*) - \mu(N^*)| =  \frac{1}{2}
\left(- (p_0 - \mu_0) + \sqrt{(p_0 - \mu_0)^2 + \frac{4 \gamma p_0}{K}}\right).
\end{align}
In Fig.~\ref{fig:Case1}, the quantity $\tilde{\lambda}_1^S$
is plotted against $\gamma$ for several different combinations of
$p_0, \mu_0$, showing the unboundedness of the convergence rate as
$\gamma$ increases.  Within this physiological range of $\gamma$
values, the dependence of $\tilde{\lambda}_1^S$ on $\gamma$ presents
as linear on a log-log plot, indicating a power law relationship.
Indeed, the power law is described in detail in the caption of
Fig.~\ref{fig:Case1}.

\subsection{Constant Proliferation, Regulated Death}
\label{sec:constprovaryd}

Let us now assume that $p(N) = p_0 > 0$ and $\mu(N) = \mu_0 +
\frac{\mu_1 N^2}{K^2 + N^2}$, with $\mu_0, \mu_1 > 0$, in the
determination of $N(t)$ via Eq.~\ref{eq:N!}. We assume that $p_0 >
\mu_0$ and $p_0 - (\mu_0 + \mu_1) < 0$, so that the action of the
proliferation-death mechanism results in net cellular birth at low
cell counts and net cellular death at high cell counts.
%
%
%
The steady states of Eq.~\ref{eq:N!} are given by the roots of the following
cubic

\begin{align} \label{eq:CASE2NROOT}
P(N) = \left(p_0 - (\mu_0 + \mu_1)\right) N^3 +
\gamma_0 N^2 + (p_0 - \mu_0) K^2 N + \gamma K^2.
\end{align}
First note that $P(0) = \gamma K^2 > 0$, and the highest order
coefficient satisfies $\left(p_0 - (\mu_0 + \mu_1)\right) < 0$ by
assumption, so that $P(N) \longrightarrow - \infty$ as $N
\longrightarrow \infty$, and $P(N)$ has at least one positive, real
root.  From Descartes' rules of signs, the polynomial has at most one
positive real root, so we may conclude that it has precisely one
positive real root.  This root corresponds to the only physically
relevant stable fixed point of $\frac{{\textrm d}N}{{\textrm d}t}$.
By regarding this root, $N^*$, as the intersection of the line $\gamma
+ (p_0 - \mu_0) N$ and the rational expression $\mu_1
\left(\frac{N^3}{K^2 + N^2}\right)$, we see that $N^* \to \infty$ as
$\gamma \to 0$.

\begin{figure}[h]
\centering
\includegraphics[width=6.5in]{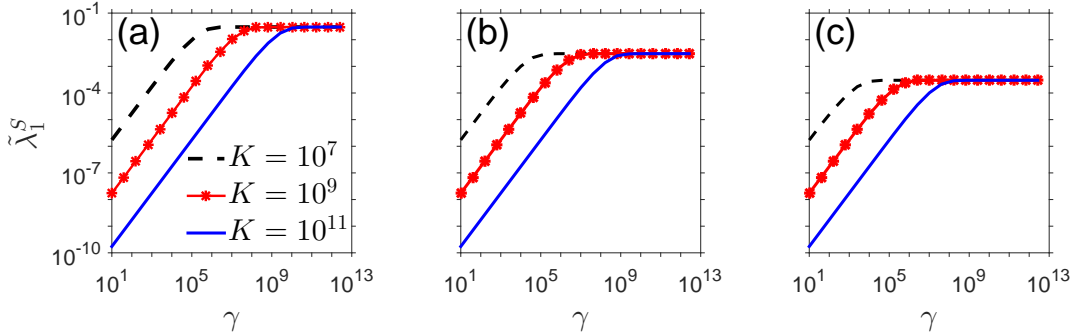}
    \caption[Dominant eigenvalue, $\tilde{\lambda}_1^S$, of
      $L_{\tilde{S}}$, plotted against $\gamma$, Case
      2]{\textbf{Dominant eigenvalue, $\tilde{\lambda}_1^S$, of
        $L_{\tilde{S}}$, plotted against $\gamma$, Case 2.} In (a),
      $p_0 = 0.18$, $\mu_0 = 0.17$, and $\mu_1 = 0.004$.  In (b), $p_0
      = 0.018$, $\mu_0 = 0.017$, and $\mu_1 = 0.004$.  In (c), $p_0 =
      0.0018$, $\mu_0 = 0.0017$, and $\mu_1 = 0.0004$.  The
      relationship between $\tilde{\lambda}_1^S$ and $\gamma$ follows
      a power law for low values of $\gamma$, before reaching a
      plateau for high values of $\gamma$.  In (a), the best fit line
      to the curve $K = 10^7$ over the power law region ($\sim \gamma
      \in [10^{1}, 2.2 \times 10^4]$) is given by
      $\log(\tilde{\lambda}_1^S) = (1.492\times 10^{-7}) \log(\gamma)
      + 1.972 \times 10^{-5}$, with $R^2 = 0.9979$.  The curve $K =
      10^9$ over the power law region ($\sim \gamma \in [10^{1}, 2.1
        \times 10^6]$) is fitted by $\log(\tilde{\lambda}_1^S) =
      (1.505 \times 10^{-9}) \log(\gamma) + 1.039 \times 10^{-5}$,
      with $R^2 = 0.998$, and the curve $K = 10^{11}$ over the power
      law region ($\sim \gamma \in [10^{1}, 2.2 \times 10^8]$) is
      fitted by $\log(\tilde{\lambda}_1^S) = (1.502 \times 10^{-11})
      \log(\gamma) + 7.6 \times 10^{-6}$, with $R^2 = 0.998$.  }
    \label{fig:Case2}
\end{figure}

We also verify that the eigenvalues $\tilde{\lambda}^S_1,
\tilde{\lambda}^S_{M+1}$ satisfy $0 > \tilde{\lambda}^S_1 >
\tilde{\lambda}^S_{M+1}$, so that $\tilde{\lambda}^S_1$ is, in fact,
the dominant eigenvalue.  We first check that $0 >
\tilde{\lambda}^S_1$.  Recalling that $\tilde{\lambda}^S_1 = p(N^*) -
\mu(N^*) = p_0 - \left(\mu_0 + \mu_1\left(\frac{(N^*)^2}{(K^2 +
  (N^*)^2)}\right)\right)$, we see after some simple algebraic
manipulation that the condition $\tilde{\lambda}^S_1 < 0$ is
equivalent to

\begin{align}\label{eq:COND}
N^* > \left(\frac{(p_0 - \mu_0)K^2}{|p_0 -
(\mu_0 + \mu_1)|}\right)^{\frac{1}{2}} := \overline{N}.
\end{align}
But $P(\overline{N}) = \frac{\gamma (p_0 - \mu_0) K^2}{|p_0
  - (\mu_0 + \mu_1)|} + \gamma K^2 > 0$.  That $P(\overline{N}) > 0$
and $P(N) \longrightarrow - \infty$ as $N \longrightarrow \infty$,
along with the fact that $P(N)$ has only one real positive root
indicates that $N^*$ does, in fact, satisfy Cond.~\ref{eq:COND}, so
that $\tilde{\lambda}^S_1 < 0$.  It is easily verified that
$-\mu'(N^*) N^* < 0$, and consequently that $\tilde{\lambda}_1^S >
\tilde{\lambda}_{M+1}^S$.

From the fact that $N^* \to \infty$ as $\gamma \to \infty$, we have
that $|p(N^*) - \mu(N^*)| \to p_0 - (\mu_0 + \mu_1)$ as $\gamma \to
\infty$.  This limiting behavior is reflected in the eventual plateau
seen in Fig.~\ref{fig:Case2}, which plots the quantity
$\tilde{\lambda}_1^S$ in this case.  Before the plateau occurs,
$\gamma$ and $\tilde{\lambda}_1^S$ are again related by a power law.
The transition from power law to plateau occurs at a ``threshold"
value, $\gamma^*$, of $\gamma$, at which the rate of T cell adjustment
becomes sensitive to a changing thymic export rate.  If, for some $i$,
$\gamma_{i-1}, \gamma_i \geq \gamma^*$, $\tilde{\lambda}_1^S$ is
unaffected by the transition from thymic export rate $\gamma_{i-1}$ to
thymic export rate $\gamma_i$--that is, the T cell pool adjusts to the
new thymic export rate $\gamma_i$ as quickly as it had adjusted to the
previous thymic export rate $\gamma_{i-1}$.  If, however, $(\gamma_i -
\gamma^*)(\gamma_{i-1} - \gamma^*) < 0$, then a dramatic shift in the
adjustment rate will occur.  Thus, parameter choices that result in a
low threshold value $\gamma^*$ might correspond to physiological
conditions under which an instance of acute thymic atrophy actually
does not affect T cell adjustment rates.  Likewise, a high threshold
value of $\gamma^*$ indicates potential sensitivity of adjustment
rates to the changing level of thymic export, with adjustment rates
obeying a power law dependence on $\gamma$.

\section{Discussion and Conclusions}
\label{sec:chapter3discussion}

In this paper, we formulated a model of how the naive T cell pool
adjusts to changes in the rate of thymic export of new T cells during
a cycle of stress-induced atrophy and recovery, and how it may be
reconstituted following an instance of severe lymphopenia induced by a
state of immune disease, or treatments such as chemotherapy.  Our
underlying model is a birth-death-immigration process studied under a
mean-field approximation for the mean clone abundance distribution (or
clone count) $c_{k}(t)$. A recent investigation into the fully
stochastic model indicates that the true $c_{k}$ differs from the
$c_{k}$ derived using the mean-field assumption
(Eqs.~\ref{eq:CLONE1}--\ref{eq:CLONEM} and \ref{eq:N}) only for very
large $k \approx N$ \cite{XU_JPA}.  Thus, our analyses may be
inaccurate only if a single large clone dominates the whole
population.

Another modeling choice we made is that TCRs are generated one naive T
cell at a time.  Successive emigrations from the thymus are
uncorrelated with the TCRs that are produced.  However, emigration can
be clustered, where there is cell proliferation of a thymocyte
generating $\Delta\sim 2-4$ copies of naive T cells carrying the same
TCR during each emigration event. In this case, we simply modify the
immigration terms in Eqs.~\ref{eq:CLONE1}-\ref{eq:CLONEk}. For
Eq.~\ref{eq:CLONE1}, the immigration term proportional to
$\gamma/\Omega$ is removed, while the immigration term
$\gamma/\Omega(c_{k-1}-c_{k})$ is replaced by
$\gamma/\Omega(c_{k-\Delta}-c_{k})$ in Eq.~\ref{eq:CLONEk}.  By
setting $c_{\ell < 0}=0$, the solution to
Eqs.~\ref{eq:CLONE1}-\ref{eq:CLONEk} can be numerically evaluated but
a closed-form analytic solution is not possible. Solutions with
clustered immigration ($\Delta > 1$) show no qualitative difference
with $\Delta = 1$, with minor quantitative differences arising only
for very small $k$.

In Section~\ref{sec:recovery}, we found that our mean-field ODE model
admitted one stable equilibrium solution when $\gamma > 0$.  From an
analysis of the eigenvalues and eigenvectors of the system linearized
around this stable equilibrium, we found that for small $k$,
perturbations in $c_k$ about a steady-state solution are
weighted more strongly in the slowest mode (slowest eigenvalue)
$\lambda_{1}^{S} = p(N^{*})-\mu(N^{*})<0$.  As also shown in
Fig.~\ref{fig:thesis3fig1}, the variation in $c_{k}$ for larger $k$
contains higher weights of the faster modes corresponding to more
negative (faster) eigenvalues $\lambda_{\ell}^{S} = \ell
(p(N^{*})-\mu(N^{*}))<0$.  Similarly, in Section~\ref{sec:atrophy}, we
analyzed the eigenvalue and eigenvector decomposition of the solution
for $\gamma=0$, for which two equilibrium points, $N^{*}=0$ and
$N^{*}>0$, arise. For $p(N^{*})-\mu(N^{*}) < 0$, we find the same
decomposition of $c_{k}(t)$ as for the $\gamma > 0$ case in
Section~\ref{sec:recovery}. Thus, for relaxation of $c_{k}(t)$ towards
a finite steady-state, our eigenvalue/eigenvector analyses suggests
that the counts of large clones might evolve faster towards
that of the new steady-state. For the unstable equilibrium
state $N^{*}=0$, the eigenvalue/eigenvector decomposition of $c_{k}$
is shown in Fig.~\ref{fig:thesis3fig3}. In this unstable case, the
slowest eigenvalue $\lambda_{0}^{U}\approx 0$ has a corresponding
eigenvector $z_{0}^{i}$ with decaying elements with $i$. This
result predicts that high-population ($M-i$) clones relax slowly
(recall that the labeling is inverted: $i=M-1$ corresponds to the
direction of $c_{1}$).


In addition to decomposing the linearized solutions in terms of
eigenvalues and eigenvectors, we explicitly plot trajectories of
$c_{k}(t)$ following small, abrupt changes in $\gamma$.
\begin{figure}[h]
\centering
\includegraphics[width=5.6in]{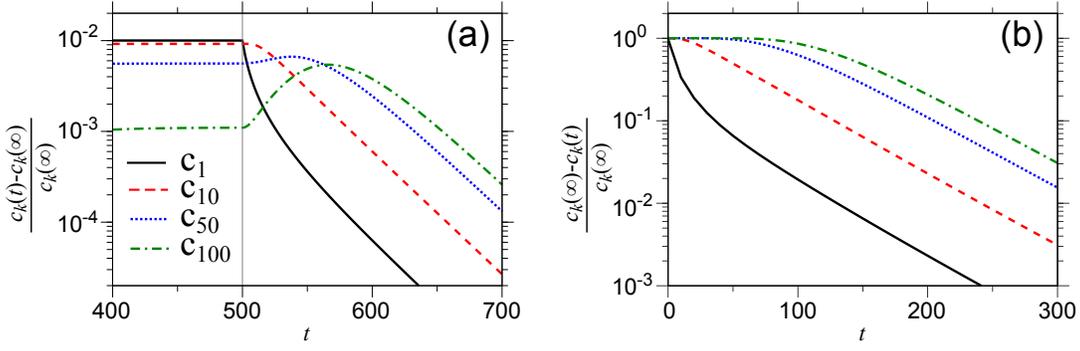}
    \caption[Time-dependence of $c_{k}(t)$ near fixed points]
            {\textbf{Time-dependence of $c_{k}(t)$ near fixed points.}
              (a) The explicit time-evolution of $\delta c_{k} =
              (c_{k}(t)-c_{k}(\infty))/c_{k}(\infty)$ following a
              small abrupt change $\gamma=1.8\times 10^{10} \to
              0.99\times 1.8\times 10^{10}$ at $t=0$. The evolution
              from one nonzero steady state to another nonzero steady
              state shows that clone counts of small clones appear to
              evolve faster than counts of larger clones.  (b)
              Similarly, the number of small clones also evolve faster
              away from an unstable empty equilibrium state
              $N^{*}=0$.}
    \label{fig:CKT}
\end{figure}
By plotting the log of the deviation $\delta c_{k} =
(c_{k}(t)-c_{k}(\infty))/c_{k}(\infty)$ in \textit{all} cases
(Fig.~\ref{fig:CKT}(a)), we can see that counts of
\textit{small} clones evolve faster after perturbation. Although this
seems to contradict the eigen decomposition of the $\gamma>0$ case,
the coefficients and eigenvector elements corresponding to larger
$\ell$ are negative (see Figs.~\ref{fig:thesis3fig1}(b) and (d)) which
convert fast decaying modes into short bursts of growth and conspire
to cancel the fast dynamics indicated by the more negative
eigenvalues. In fact, we see that the clone counts of large clones
actually evolve more slowly than the rate associated with the largest
eigenvalue. This behavior can also be heuristically understood by
noticing that under a given change in $\gamma$, the immigration term
$\gamma (c_{k-1}-c_{k})/\Omega$ is larger for smaller $k$ because
$c_{k} \sim 1/k$. Thus, for a given change in $\gamma$, the
perturbation is larger for the equations with smaller $k$ and inducing
changes in $c_{k}(t)$ that appear larger.

In section~\ref{sec:specialcases}, we infer the rate of convergence of
the counts of the smallest (but most common) clones to
equilibrium by computing the dominant eigenvalue as a function of
$\gamma>0$ for two choices of regulated functions $p(N), \mu(N)$.  In
section~\ref{sec:logisticmodel}, we test the logistic model, which
assumes a constant death rate but a population-dependent proliferation
rate.  From the explicit form of $\tilde{\lambda}_1^S$ in
Eq.~\ref{eq:logisticeval}, $p(N^*) - \mu(N^*) \to \infty$ as $\gamma
\to \infty$ for fixed values of the other parameters, which produces
the power-law relationship between $\gamma$ and $\tilde{\lambda}_1^S$
depicted in Fig.~\ref{fig:Case1}.  In section~\ref{sec:constprovaryd},
we assumed instead a constant rate of cellular proliferation with an
$N$-dependent death rate.  In this case, which differs from the
logistic formulation in that regulation is incorporated into $\mu(N)$
via a Hill-type function, $\gamma$ and $\tilde{\lambda}_1^S$ are
related by a power law for low $\gamma$, before reaching a plateau at
higher $\gamma$.

Since regulation through death is typically associated with the actual
mechanism of naive T cell survival, we expect this mechanism to be
more realistic than a population-dependent proliferation rate $p(N)$.
Thus, jumps in the thymic export rate that either cross the threshold
value of $\gamma$, or occur between two values of $\gamma$ both in the
power-law region, can be expected to produce changes in both the
equilibrium values of $c_k$ and also the convergence rates.  If a jump
in $\gamma$ occurs between two values of $\gamma$ that are both in the
plateau region, the equilibrium values shift, but the convergence
rates stay the same.  The presence of the power law region indicates
the robustness of the T cell diversity during a time of severe thymic
atrophy.  That is, the slower convergence of the T cell diversity to
equilibrium at low $\gamma$ values protects the pool from quick shifts
to the lower diversity associated with lower $\gamma$ values.

\section*{Acknowledgements}
\vspace{-3mm} This research was supported by grants from the National
Science Foundation (DMS-1814364), the Army Research Office
(W911NF-18-1-0345), and the National Institutes of Health
(R01HL146552).

\appendix
\section{Verification of recurrence relation solution}
\label{sec:APPrecursol}
Here, we verify  that Eq.~\ref{eq:RECURSOL} satisfies the
recurrence relation in Eq.~\ref{eq:RECUR}. For notational simplicity, we
assume $p/\mu \equiv p(N^{*})/\mu(N^{*})$ is evaluated at the
relevant steady state defined by $N^{*}$.
%
%
%
%
%
Inserting Eq.~\ref{eq:RECURSOL} into the right-hand-side
of Eq.~\ref{eq:RECUR}, we find

\[\scalemath{0.8}{\left[(i + (k-1))\left(\frac{p}{\mu}\right)
+ (i-(k+1)) \right] y^{i-1}_k -
\left[i-2\right]\left(\frac{p}{\mu}\right) y^{i-2}_k}\]
\[\scalemath{0.68}{=
\left[ \frac{(i + (k-1))}{i} \left(\frac{p}{\mu}\right)
+ \frac{i - (k+1)}{i}\right] \left[
\frac{\left(\frac{p}{\mu}\right)^{i-2}}{k!} \prod_{j=1}^{k-1} (i-1+j)
+ \sum_{n=2}^{k-1}\left[\prod_{m=1}^{n-1} (i-1-m) \right]\left[\prod_{j=1}^{k-n} (i-1+j) \right]
\frac{\left(\frac{p}{\mu}\right)^{i-1-n}}{(-1)^{n-1} (n-1)! k (k-n)!}
+ \frac{\left(\frac{p}{\mu}\right)^{i-1-k}}{(-1)^{k-1} k!}
\prod_{j=1}^{k-1} (i-1-j) \right]}\]
\[\scalemath{0.72}{- \left(\frac{p}{\mu}\right)
\left(\frac{i-2}{i}\right) \left[\left(\prod_{j=1}^{k-1} (i-2-j)\right)
\frac{\left(\frac{p}{\mu}\right)}{k!} + \sum_{n=2}^{k-1}
\left[\prod_{m=1}^{n-1}(i-2-m)\right]\left[\prod_{j=1}^{k-n}(i-2+j)\right]
\frac{\left(\frac{p}{\mu}\right)}{(-1)^{n-1} (n-1)! k (k-n)!} +
\frac{\left(\frac{p}{\mu}\right)^{i-2-k}}{(-1)^{k-1} k!}
\prod_{j=1}^{k-1} (i-2-j)\right]}\]

\vskip12pt

\[\scalemath{0.75}{= \left[\prod_{j=0}^{k-1} (i+j)\right]
\frac{\left(\frac{p}{\mu}\right)^{i-1}}{k!} +
\left[\frac{(i-(k+1))}{k!}\left[\prod_{j=1}^{k-1} (i-1+j)\right]
- \frac{(i+(k-1))}{k (k-2)!} \left[\prod_{j=1}^{k-2} (i-1+j)\right]
- \frac{(i-2)}{k!} \left[\prod_{j=1}^{k-1} (i-2+j)\right]\right]
\left(\frac{p}{\mu}\right)^{i-2}}\]
\[\scalemath{0.7}{+ \sum_{n=2}^{k-1} (i-(k+1))\left[\prod_{j=1}^{n-1} (i-1-j)\right]
\left[\prod_{j=1}^{k-n} (i-1+j)\right]
\frac{\left(\frac{p}{\mu}\right)^{i-1-n}}{(-1)^{n-1} (n-1)! k (k-n)!} +
\sum_{n=3}^{k-1} (i+(k-1))\left[\prod_{j=1}^{n-1} (i-1-j)\right]
\left[\prod_{j=1}^{k-n} (i-1+j)\right]
\frac{\left(\frac{p}{\mu}\right)^{i-n}}{(-1)^{n-1} (n-1)! k (k-n)!}}\]
\[\scalemath{0.7}{+ (i+(k-1))\left[\prod_{j=1}^{k-1} (i-1-j)\right]
\left[\frac{\left(\frac{p}{\mu}\right)^{i-k}}{(-1)^{k-1} k!}\right]
+ (i-(k+1))\left[\prod_{j=1}^{k-1} (i-1-j)\right]
\frac{\left(\frac{p}{\mu}\right)^{i-1-k}}{(-1)^{k-1} k!}}\]
\[\scalemath{0.7}{-\sum_{n=2}^{k-1} (i-2)
\left[\prod_{m=1}^{n-1} (i-2-m)\right] \left[\prod_{j=1}^{k-n} (i-2+j)\right]
\frac{\left(\frac{p}{\mu}\right)^{i-1-n}}{(-1)^{n-1} (n-1)! k (k-n)!}
- (i-2) \left[\prod_{j=1}^{k-1} (i-2-j)\right]
\frac{\left(\frac{p}{\mu}\right)^{i-1-k}}{(-1)^{k-1} k!}}\]

\vskip12pt

\[\scalemath{0.72}{= \frac{\left(\frac{p}{\mu}\right)^{i-1}}{k!}\prod_{j=0}^{k-1} (i+j)+
\left(\frac{p}{\mu}\right)^{i-2}
\left[ \frac{(i+(k-2))(i-(k+1)) - (k-1)(i-2)(i+(k-1)) - (i-2)(i-1)}{k!}\right]
\prod_{j=0}^{k-3} (i+j)}\]
\[\scalemath{0.52}{+ \sum_{s=3}^{k-1} \left(\left[\prod_{j=1}^{s-1} (i-1-j)\right]
\left[\prod_{j=1}^{k-s} (i-1+j)\right] \frac{(i+(k-1))}{(-1)^{s-1} (s-1)! k (k-s)!}
+ \left[\prod_{j=1}^{s-2} (i-1-j)\right]\left[\prod_{j=1}^{k-(s-1)} (i-1+j)\right]
\frac{(i-(k+1))}{(-1)^{s-2} (s-2)! k (k-(s-1))!} - (i-2)\left[\prod_{j=1}^{s-2}(i-2-j)\right]
\left[ \prod_{j=1}^{k-(s-1)} (i-2+j)\right] \frac{1}{(-1)^{s-2} (s-2)! k (k-(s-1))!}\right)
\left(\frac{p}{\mu}\right)^{i-s}}\]
\[\scalemath{0.7}{+ \left((i-(k+1))\left[\prod_{j=1}^{k-2} (i-1-j)\right]
\frac{i}{(-1)^{k-2} (k-2)! k} + (i+(k-1))
\left[\prod_{j=1}^{k-1} (i-1-j)\right] \frac{1}{k!} - (i-2)
\left[ \prod_{j=1}^{k-2} (i-2-j)\right] \frac{(i-1)}{(-1)^{k-2} (k-2)! k}\right)
\left(\frac{p}{\mu}\right)^{i-k}}\]

\vskip12pt

\[\scalemath{0.7}{= \left[\prod_{j=0}^{k-1} (i+j)\right]
\frac{\left(\frac{p}{\mu}\right)^{i-1}}{k!} +
\left[\prod_{j=0}^{k-3} (i+j)\right] \left[\frac{-(k-1) i^2
- (k-1)(k-3) i + (k-2) (k-1)}{k!}\right] \left(\frac{p}{\mu}\right)^{i-2}}\]
\[\scalemath{0.7}{+ \sum_{s=3}^{k-1} (-1)^{-s} \left[\prod_{j=2}^{s-1} (i-j) \right]
\left[\prod_{j=0}^{k-(s+1)} (i+j)\right] \left[\frac{-(i-s)(i+(k-1))}{(s-1)! k (k-s)!}
+ \frac{(i-(k+1))(i+(k-s))}{(s-2)! k (k-(s-1))!} -
\frac{(i-1)(i-s)}{(s-2)! k (k - (s-1))!}\right]\left(\frac{p}{\mu}\right)^{i-s}}\]
\[\scalemath{0.7}{+ \left(\frac{(i+(k-1))}{(-1)^{k-1} k!}\left[\prod_{j=1}^{k-1} (i-1-j)\right]
 + \frac{(i-(k+1)) \left[\prod_{j=1}^{k-2} (i-1-j)\right] i}{(-1)^{k-2} (k-2)! k}
- \frac{(i-2)(i-1) \left[ \prod_{j=1}^{k-2} (i-2-j)\right]}{(-1)^{k-2} (k-2)! k}\right)
\left(\frac{p}{\mu}\right)^{i-k}}\]

\vskip12pt

\[\scalemath{0.7}{= \left[\prod_{j=0}^{k-1} (i+j)\right] \frac{\left(\frac{p}{\mu}\right)^{i-1}}{k!}
+ \left[\prod_{j=0}^{k-3}(i+j)\right] \left[\frac{-i^2 - (k-3) i + (k-2)}{k (k-2)!}\right] \left(\frac{p}{\mu}\right)^{i-2}}\]
\[\scalemath{0.7}{+ \sum_{s=3}^{k-1} (-1)^{-s}\left[\prod_{j=2}^{s-1} (i-j)\right]\left[\prod_{j=0}^{k-(s+1)} (i+j)\right] \left(\frac{p}{\mu}\right)^{i-s} \left[\frac{-(k-(s-1))(i-s)(i+(k-1)) + (s-1)(i-(k+1))(i+(k-s)) - (s-1)(i-1)(i-s)}{(s-1)! k (k-(s-1))!}\right]}\]
\[\scalemath{0.7}{+ \left(\frac{p}{\mu}\right)^{i-k} (-1)^{-k}
\left(\prod_{j=2}^{k-1}(i-j)\right)
\left[\frac{-(i+(k-1))(i-k) + (k-1)(i-(k+1))i - (i-1)(i-k)(k-1)}{k!}\right]}\]

\vskip12pt

\[\scalemath{0.7}{= \left[\prod_{j=0}^{k-1} (i+j)\right]
\frac{\left(\frac{p}{\mu}\right)^{i-1}}{k!} - \left[\prod_{j=0}^{k-3} (i+j)\right]
\left[\frac{(i+(k-2))(i-1)}{k (k-2)!}\right] \left(\frac{p}{\mu}\right)^{i-2}
+ (-1)^k \left(\prod_{j=2}^{k-1} (i-j)\right)\left[\frac{-i^2 + i}{k!}\right]
\left(\frac{p}{\mu}\right)^{i-k}}\]
\[\scalemath{0.7}{+ \sum_{s=3}^{k-1} \left[\prod_{j=2}^{s-1} (i-j)\right]
\left[\prod_{j=0}^{k-(s+1)}(i+j)\right] (-1)^{-s} \left(\frac{p}{\mu}\right)^{i-s}
\left[\frac{-(k-(s-1)) i^2 - (k-(s-1))(k-(s+1)) i + (k-s)(k-s+1)}{(s-1)! k (k-(s-1))!}\right]}\]

\vskip12pt

\[\scalemath{0.7}{= \left[\prod_{j=0}^{k-1} (i+j)\right]
\frac{\left(\frac{p}{\mu}\right)^{i-1}}{k!} - \left[\prod_{j=0}^{k-2} (i+j)\right]
\frac{(i-1)}{k (k-2)!} \left(\frac{p}{\mu}\right)^{i-2} +
(-1)^{-(k-1)}\left(\prod_{j=2}^{k-1} (i-j)\right) \frac{i(i-1)}{k!}
\left(\frac{p}{\mu}\right)^{i-k}}\]
\[\scalemath{0.7}{+ \sum_{s=3}^{k-1} \left[\prod_{j=2}^{s-1}(i-j)\right]
\left[\prod_{j=0}^{k-(s+1)} (i+j)\right]
\left[\frac{-i^2 - (k-(s+1)) i + (k-s)}{(s-1)! k (k-s)!}\right]
\frac{\left(\frac{p}{\mu}\right)^{i-s}}{(-1)^s}}\]

\vskip12pt

\[\scalemath{0.7}{= \left[\prod_{j=0}^{k-1} (i+j)\right]
\frac{\left(\frac{p}{\mu}\right)^{i-1}}{k!} - \left[\prod_{j=0}^{k-2} (i+j)\right]
\frac{(i-1)}{k (k-2)!}\left(\frac{p}{\mu}\right)^{i-2} + (-1)^{-(k-1)}
\left(\prod_{j=0}^{k-1} (i-j)\right) \frac{\left(\frac{p}{\mu}\right)^{i-k}}{k!}}\]
\[\scalemath{0.7}{+ \sum_{s=3}^{k-1} \left[\prod_{j=1}^{s-1} (i-j)\right] \left[\prod_{j=0}^{k-s} (i+j)\right] \frac{\left(\frac{p}{\mu}\right)^{i-s}}{(-1)^{s-1} (s-1)! k (k-s)!}}\]

\vskip12pt

\[\scalemath{0.7}{= \left[\prod_{j=0}^{k-1} (i+j)\right]
\frac{\left(\frac{p}{\mu}\right)^{i-1}}{k!} + \sum_{s=2}^{k-1}
\left[\prod_{j=1}^{s-1} (i-j)\right] \left[\prod_{j=0}^{k-s} (i+j)\right]
\frac{\left(\frac{p}{\mu}\right)^{i-s}}{(-1)^{s-1} (s-1)! k (k-s)!} +
\left[\prod_{j=0}^{k-1} (i-j)\right] \frac{\left(\frac{p}{\mu}\right)^{i-k}}{(-1)^{k-1} k!}}\]

\vskip12pt

\[\scalemath{0.7}{= i\left[\left[\prod_{j=1}^{k-1} (i+j)\right]
\frac{\left(\frac{p}{\mu}\right)^{i-1}}{k!} +
\sum_{s=2}^{k-1} \left[\prod_{j=1}^{s-1}(i-j)\right]
\left[\prod_{j=1}^{k-s} (i+j)\right]
\frac{\left(\frac{p}{\mu}\right)^{i-s}}{(-1)^{s-1} (s-1)! k (k-s)!}
+ \left[\prod_{j=1}^{k-1} (i-j)\right]
\frac{\left(\frac{p}{\mu}\right)^{i-k}}{(-1)^{k-1} k!}\right] = i y_k^i}\]


\bibliographystyle{unsrt}
\bibliography{atrophybib2}

\end{document}